\documentclass{aa}

\usepackage[T1]{fontenc}
\usepackage[varg]{txfonts}
\usepackage{mathptmx}
\usepackage{amsmath}
\usepackage{amssymb}
\usepackage{amstext}
\usepackage{amsfonts}
\usepackage{mathrsfs}
\usepackage{wasysym}

\usepackage{graphicx}

\bibpunct{(}{)}{;}{a}{}{,}
\newcommand{\mockalph}[1]{}
\defcitealias{Schafer2020}{SJB20}

\usepackage{pifont}
\newcommand{\cmark}{\ding{51}}
\newcommand{\xmark}{\ding{55}}

\newcommand{\dust}{\textrm{d}}
\newcommand{\gas}{\textrm{g}}
\newcommand{\St}{\textrm{St}}

\newcommand{\mm}{\textrm{mm}}
\newcommand{\cm}{\textrm{cm}}
\newcommand{\au}{\textrm{au}}
\newcommand{\yr}{\textrm{yr}}
\newcommand{\kyr}{\textrm{kyr}}
\newcommand{\g}{\textrm{g}}
\newcommand{\K}{\textrm{K}}

\begin{document}

\title{The coexistence of the streaming instability and the vertical shear instability in protoplanetary disks}
\subtitle{Planetesimal formation thresholds explored in two-dimensional global models}
\author{Urs Sch{\"a}fer\inst{\ref{Copenhagen}}
\and Anders Johansen{\inst{\ref{Copenhagen},\ref{Lund}}}}
\institute{Centre for Star and Planet Formation, Globe Institute, University of Copenhagen, {\O}ster Voldgade 5-7, 1350 Copenhagen, Denmark\\\email{urs.schafer@sund.ku.dk}\label{Copenhagen}
\and Lund Observatory, Department of Astronomy and Theoretical Physics, Lund University, Box 43, 22100 Lund, Sweden\label{Lund}}
\date{}
\abstract{The streaming instability is a promising mechanism to induce the formation of planetesimals. Nonetheless, this process has been found in previous studies to require either a dust-to-gas surface density ratio or a dust size that is enhanced compared to observed values. Employing two-dimensional global simulations of protoplanetary disks, we show that the vertical shear instability and the streaming instability in concert can cause dust concentration that is sufficient for planetesimal formation for lower surface density ratios and smaller dust sizes than the streaming instability in isolation, and in particular under conditions that are consistent with observational constraints. This is because dust overdensities forming in pressure bumps induced by the vertical shear instability act as seeds for the streaming instability and are enhanced by it. While our two-dimensional model does not include self-gravity, we find that strong dust clumping and the formation (and dissolution) of gravitationally unstable overdensities can be robustly inferred from the evolution of the maximum or the mean dust-to-gas volume density ratio. The vertical shear instability puffs up the dust layer to an average mid-plane dust-to-gas density ratio that is significantly below unity. We therefore find that reaching a mid-plane density ratio of one is not necessary to trigger planetesimal formation via the streaming instability when it acts in unison with the vertical shear instability.}
\titlerunning{Thresholds for planetesimal formation owing to streaming instability and vertical shear instability}
\authorrunning{U. Sch{\"a}fer and A. Johansen}
\maketitle

\section{Introduction}
\label{sect:introduction}
The streaming instability has emerged in the last decade as the leading candidate among processes to cause the formation of planetesimals \citep{Chiang2010, Johansen2014}, a key step in the growth from sub-micron-sized dust grains to planets. The streaming instability was discovered analytically as a linear instability, that is to say an exponential growth of small linear perturbations, by \citet{Youdin2005}. Subsequently, it was shown how the instability evolves from a linear into a non-linear phase in numerical simulations \citep{Johansen2007a, Bai2010a, Kowalik2013, Mignone2019}. These simulations reveal that the instability in its non-linear regime causes dust to accumulate in largely axisymmetric filaments \citep{Johansen2007a, Bai2010b, Kowalik2013, Yang2014, Li2018}. Overdensities in these filaments can be sufficiently large to contract owing to their self-gravity and form planetesimals with typical sizes of tens or hundreds of kilometres \citep{Johansen2007b, Johansen2009, Johansen2015, Simon2016, Schafer2017, Abod2019}. 

In all likelihood, the streaming instability is active everywhere dust is present in protoplanetary disks. The instability requires only rotating gas and dust which are coupled via mutual drag, and a gas pressure gradient with respect to the radial distance to the star \citep{Youdin2005}. This gradient entails a slight deviation of the orbital speed of the gas and dust components of protoplanetary disks from the Keplerian speed as well as a radial drift of the two components, especially a drift of the dust towards the star \citep[e.g.][]{Adachi1976, Nakagawa1986, Brauer2007}. Local dust overdensities drift more slowly than the surrounding dust, resulting in further dust accumulation in the overdensities, a further reduction of their drift speed, and ultimately instability -- the streaming instability.

However, the ability of the streaming instability to give rise to dust concentration that is strong enough to trigger planetesimal formation depends on three parameters: the Stokes number of the dust~$\St$, the dust-to-gas surface density ratio~$\Sigma_{\dust}/\Sigma_{\gas}$\footnote{Although the dust-to-gas surface density ratio is often referred to as metallicity~$Z$ in this context, we avoid this term here. This is because the term is more generally used to denote the abundance of elements heavier than helium, which is related but not equivalent to the dust-to-gas ratio.}, and the strength of the radial gas pressure gradient \citep{Johansen2009, Bai2010b, Bai2010c, Drazkowska2014, Carrera2015, Yang2017, Li2021}. The latter is commonly expressed in terms of the dimensionless parameter~$\Pi$ as introduced by \citet{Bai2010b}, which is typically of the order of~$0.01$ or~$0.1$. \citet{Sekiya2018} propose, and corroborate using simulations, a dependence on the parameter~$(2\pi)^{1/2}\,\Sigma_{\dust}/(\Sigma_{\gas}\Pi)$ rather than on~$\Sigma_{\dust}/\Sigma_{\gas}$ and~$\Pi$ individually. This parameter expresses the dust-to-gas density ratio in the dust mid-plane layer under the assumption that the thickness of this layer is regulated by the streaming instability. \citet{Carrera2015}, \citet{Yang2017}, and \citet{Li2021} each conducted a suite of simulations to obtain, for a given Stokes number and pressure gradient, a minimum dust-to-gas surface density ratio necessary for planetesimal formation. Ever-smaller minimum values are found in these studies, with \citet{Li2021} inferring a value of~\mbox{$\Sigma_{\dust}/\Sigma_{\gas}=2\%$} for~\mbox{$\St=0.01$} and~\mbox{$\Pi=0.05$} as well as~\mbox{$\Sigma_{\dust}/\Sigma_{\gas}=0.6\%$} for the same pressure gradient but~\mbox{$\St=0.1$}.

These numerical results raise the question of whether the streaming instability on its own can be expected to cause planetesimal formation for the dust sizes and dust-to-gas ratios that are observed in protoplanetary disks. We focus here on Class II disks as we numerically model such disks in this study. Notably, ALMA has enabled measurements of dust sizes and dust-to-gas ratios in these disks in recent years owing to its high spatial resolution and sensitivity.

Maximum dust sizes are most frequently inferred from either the opacity spectral index of thermal dust emission \citep[e.g.][]{Draine2006, Testi2014} or from the polarisation of the dust emission owing to self-scattering by other dust grains or grain alignment \citep{Kataoka2015, Kataoka2017, Mori2021a}. The former method yields millimetre- or centimetre-sizes, with a trend towards smaller sizes at greater distances to the star \citep{Perez2012, Perez2015, Tazzari2016, Tazzari2021a, Tazzari2021b, Sierra2019, Macias2019, Macias2021,  Carrasco-Gonzalez2019, Mauco2021}. In contrast, sizes of not more than a few hundreds of microns are derived applying the latter method \citep{Kataoka2017, Bacciotti2018, Ohashi2019, Ohashi2020, Lin2020, Mori2021a}.

To explain the inconsistency between these two kinds of measurements, it has been proposed that the sizes obtained from spectral indices might be overestimates that stem from optically thick emission being misinterpreted as optically thin, in particular if self-scattering is neglected \citep{Liu2019, Zhu2019, Lin2020, Ohashi2020}. However, this degeneracy can be broken when multi-wavelength observations are considered, with the sizes inferred from modeling such observations still lying in the millimetre-to-centimetre range (\citealt{Sierra2019, Carrasco-Gonzalez2019, Macias2021, Mauco2021}, see also \citealt{Tapia2019})\footnote{Nonetheless, additional degeneracy arises from the dust size distribution \citep{Sierra2019, Macias2021}.}. On the other hand, the deviations from spectral index measurements can be alleviated when the shape and chemical composition of dust grains are taken into account in polarisation measurements \citep{Kirchschlager2020, Yang2020, Brunngraber2021}.

The canonical dust-to-gas ratio in the Milky Way interstellar medium amounts to~${\sim}1\%$. The ratio is not constant, however, but varies by factors of a few even within our galaxy \citep{Spitzer1978, Sodroski1997} -- particularly also in molecular clouds \citep{Liseau2015} -- and by orders of magnitude when considering galaxies with different metallicities \citep{Brinchmann2013, RemyRuyer2014}. The ratio of dust to gas mass in protoplanetary disks is observed to typically be higher than the canonical interstellar medium value, with ratios of~${\sim}10\%$ being common \citep{Ansdell2016, Miotello2017, Long2017, Wu2018, Macias2021}. Nevertheless, measurements of both dust and gas masses are subject to significant uncertainties. On the one hand, the total dust mass can be reliably inferred from the thermal dust emission only if the emission is optically thin \citep[e.g.][]{Andrews2020}. On the other hand,~H$_2$ -- the most abundant gas molecule -- is difficult to detect, conversion from the observed mass of~CO -- the second most abundant molecule -- to the total gas mass is complex and prone to error, and observations of the promising tracer molecule~HD are still rare \citep[e.g.][]{Miotello2017, Andrews2020, Anderson2022}. Based on the composition of the solar photosphere as well as meteorites, \citet{Lodders2003} finds that condensates should have constituted~${\sim}1.5\%$ of the mass of the protoplanetary disk which evolved into the Solar System. 

Comparing these estimates of the dust sizes and dust-to-gas ratios in protoplanetary disks with the results of the above-mentioned parameter studies \citep{Carrera2015, Yang2017, Li2021} shows that planetesimal formation induced by the streaming instability alone is possible at around or slightly higher than the canonical interstellar medium dust-to-gas ratio. In this context, a variety of mechanisms have been proposed to act in concert with the streaming instability. Among them are two other hydrodynamical instabilities, the subcritical baroclinic instability \citep{Raettig2015, Raettig2021} and the vertical shear instability \citep[][hereafter SJB20]{Lehmann2022, Schafer2020}. 

The vertical shear instability is a hydrodynamical instability that arises if the gas orbital velocity varies with height -- for instance owing to a radial temperature gradient, which is omnipresent in protoplanetary disks \citep[e.g.][]{Andrews2020} -- and the gas cooling timescale is sufficiently short \citep{Nelson2013, Lin2015}. \citet{Stoll2016} find that the vertical shear instability gives rise to transient pressure fluctuations in which dust accumulates, resulting in overdensities of several times the mean initial dust density. Similarly, dust is concentrated in vortices formed by the instability \citep{Flock2017, Flock2020, Lehmann2022}.

In a previous paper \citepalias{Schafer2020}, we present evidence that overdensities induced by the vertical shear instability trigger further concentration by the streaming instability. The aim of this paper is to examine to what extent a combination of the vertical shear instability and the streaming instability leads to a reduction of the minimum values of dust-to-gas surface density ratio and dust size or Stokes number required for planetesimal formation compared to if only the streaming instability is considered. To this end, we performed a parameter study adopting the numerical model of \citetalias{Schafer2020}. Our two-dimensional, axisymmetric simulations cover the full scale of protoplanetary disks, while adaptive mesh refinement permits us to locally resolve the formation of dust overdensities in the mid-plane layer owing to the two instabilities.

In Section~\ref{sect:simulations}, we describe our numerical model. We examine dust concentration and the relation between dust overdensities and pressure bumps in Sect.~\ref{sect:dust_concentration}. In Section~\ref{sect:metrics}, we analyse which metrics can be applied to gauge the potential for planetesimal formation in models such as ours that do not include self-gravity. This is followed by a presentation of the threshold values of dust-to-gas surface density ratio and dust size that are necessary for the streaming instability alone or it and the vertical shear instability in combination to induce dust concentration that is sufficient for planetesimal formation in Sect.~\ref{sect:thresholds}. We discuss the implications and limitations of our study in Sect.~\ref{sect:discussion} and summarise its results in Sect.~\ref{sect:summary}.

\section{Simulations}
\label{sect:simulations}
\begin{table*}
\caption{Simulation parameters}
\centering
\resizebox{\hsize}{!}{
\begin{tabular}{lccccccc}
\hline
\hline
Simulation name&Equation&Drag of dust&$t_{\textrm{d,init}}$ [$\kyr$]\tablefootmark{a}&$L_z$ [$H_{\gas}$]\tablefootmark{b}&$\Sigma_{\dust}/\Sigma_{\gas}$ [\%]\tablefootmark{c}&Dust par-&$t_{\textrm{end}}$ [$\kyr$]\tablefootmark{e}\\
&of state&onto gas?&&&&ticle size\tablefootmark{d}&\\
\hline
\hline
\textit{VSI}&isothermal&no&$50$&$4$&$2$&\mbox{$a=3~\cm$}&$55$\\
\hline
\textit{SI\_0.5\_St=0.1}&adiabatic&yes&$0$&$2$&$0.5$&\mbox{$\St=0.1$}&$2.5$\\
\textit{SI\_0.5\_a=3}&adiabatic&yes&$0$&$2$&$0.5$&\mbox{$a=3~\cm$}&$10$\\
\textit{SI\_1\_St=0.1}&adiabatic&yes&$0$&$2$&$1$&\mbox{$\St=0.1$}&$10$/$10$\tablefootmark{f}\\
\textit{SI\_1\_a=3}&adiabatic&yes&$0$&$2$&$1$&\mbox{$a=3~\cm$}&$10$\\
\textit{SI\_2\_a=0.3}&adiabatic&yes&$0$&$2$&$2$&\mbox{$a=3~\mm$}&$10$\\
\textit{SI\_2\_St=0.1}&adiabatic&yes&$0$&$2$&$2$&\mbox{$\St=0.1$}&$2.5$\\
\textit{SI\_2\_a=3}&adiabatic&yes&$0$&$2$&$2$&\mbox{$a=3~\cm$}&$2.5$\\
\hline
\textit{SIwhileVSI\_0.5\_St=0.1}&isothermal&yes&$0$&$4$&$0.5$&\mbox{$\St=0.1$}&$5$\\
\textit{SIwhileVSI\_0.5\_a=3}&isothermal&yes&$0$&$4$&$0.5$&\mbox{$a=3~\cm$}&$10$\\
\textit{SIwhileVSI\_1\_St=0.1}&isothermal&yes&$0$&$4$&$1$&\mbox{$\St=0.1$}&$10$\\
\textit{SIwhileVSI\_1\_a=3}&isothermal&yes&$0$&$4$&$1$&\mbox{$a=3~\cm$}&$10$\\
\textit{SIwhileVSI\_2\_a=0.3}&isothermal&yes&$0$&$4$&$2$&\mbox{$a=3~\mm$}&$10$\\
\textit{SIwhileVSI\_2\_St=0.1}&isothermal&yes&$0$&$4$&$2$&\mbox{$\St=0.1$}&$5$\\
\textit{SIwhileVSI\_2\_a=3}&isothermal&yes&$0$&$4$&$2$&\mbox{$a=3~\cm$}&$10$\\
\hline
\textit{SIafterVSI\_0.5\_St=0.01}&isothermal&yes&$50$&$4$&$0.5$&\mbox{$\St=0.01$}&$55$\\
\textit{SIafterVSI\_0.5\_a=0.3}&isothermal&yes&$50$&$4$&$0.5$&\mbox{$a=3~\mm$}&$60$\\
\textit{SIafterVSI\_0.5\_St=0.1}&isothermal&yes&$50$&$4$&$0.5$&\mbox{$\St=0.1$}&$55$\\
\textit{SIafterVSI\_0.5\_a=3}&isothermal&yes&$50$&$4$&$0.5$&\mbox{$a=3~\cm$}&$55$\\
\textit{SIafterVSI\_1\_St=0.01}&isothermal&yes&$50$&$4$&$1$&\mbox{$\St=0.01$}&$55$\\
\textit{SIafterVSI\_1\_a=0.3}&isothermal&yes&$50$&$4$&$1$&\mbox{$a=3~\mm$}&$60$/$55$\tablefootmark{f}\\
\textit{SIafterVSI\_1\_St=0.1}&isothermal&yes&$50$&$4$&$1$&\mbox{$\St=0.1$}&$55$\\
\textit{SIafterVSI\_1\_a=3}&isothermal&yes&$50$&$4$&$1$&\mbox{$a=3~\cm$}&$55$\\
\textit{SIafterVSI\_2\_St=0.01}&isothermal&yes&$50$&$4$&$2$&\mbox{$\St=0.01$}&$60$\\
\textit{SIafterVSI\_2\_a=0.3}&isothermal&yes&$50$&$4$&$2$&\mbox{$a=3~\mm$}&$55$\\
\textit{SIafterVSI\_2\_St=0.1}&isothermal&yes&$50$&$4$&$2$&\mbox{$\St=0.1$}&$52$\\
\textit{SIafterVSI\_2\_a=3}&isothermal&yes&$50$&$4$&$2$&\mbox{$a=3~\cm$}&$60$\\
\hline
\hline
\end{tabular}
}
\tablefoot{
\tablefoottext{a}{Time after which particles representing the dust are initialised.}
\tablefoottext{b}{Vertical domain size, where~$H_{\rm g}$ is the gas scale height.}
\tablefoottext{c}{Dust-to-gas surface density ratio.}
\tablefoottext{d}{Given either as a size~$a$ or as a Stokes number~$\St$.}
\tablefoottext{e}{Time after which simulation ends.}
\tablefoottext{f}{Simulation with doubled resolution.}
}
\label{table:simulations}
\end{table*}

We applied the FLASH Code\footnote{http://flash.uchicago.edu/site/flashcode/}\textsuperscript{,}\footnote{While we are not permitted to re-distribute the FLASH Code or any of its
parts, we are happy to share the modifications to the
code that we implemented to perform the simulations presented
in \citetalias{Schafer2020} and in this paper.} \citep{Fryxell2000} to conduct simulations of the gas and dust components of protoplanetary disks, including the mutual drag between the two components and the stellar gravity. Gas and dust were modeled on a Eulerian grid and as Lagrangian particles, respectively. We summarise in this section the most important aspects of the simulations, and refer to \citetalias{Schafer2020} for a more detailed description.

Table~\ref{table:simulations} lists all simulations and the parameters that distinguish them. The simulations names are composed of, in this order, the instabilities which are simulated, the initial dust-to-gas surface density ratio, as well as the dust size or Stokes number. We discern two kinds of simulations in which both the streaming instability and the vertical shear instability are active: In the scenario \emph{SIwhileVSI}, both instabilities start to grow at the same time. On the other hand, the vertical shear instability has already saturated before the streaming instability begins to operate in the scenario \emph{SIafterVSI}. The latter scenario was realised by initialising the dust particles after~$50~\kyr$ instead of at the beginning of the simulations. Furthermore, to exclude the streaming instability and simulate only the vertical shear instability we neglected the drag exerted by the dust onto the gas. 

\subsection{Domain size and resolution}
The geometry of our two-dimensional simulation domains is cylindrical as this is a natural choice to model protoplanetary disks. The axisymmetric domains range from~$10$ to~$100~\au$ in the radial dimension while encompassing~$1$ or~$2$ gas scale heights above and below the disk mid-plane. All simulations involving the vertical shear instability were performed using the larger vertical domain size of~$4$ scale heights since in \citetalias{Schafer2020} we show that this size is required to reproduce the turbulent strength which is found in previous numerical studies of this instability. Gas and dust were permitted to leave but not enter the domains through the radial and vertical boundaries, with the pressure being interpolated to ensure vertical hydrostatic equilibrium at the latter boundaries. To avoid that artificial behaviour caused by the inner radial boundary conditions contaminates our results, we excluded the innermost~$10~\au$ from all quantitative analysis. (The dust rapidly drifting and settling away from them renders a similar treatment of outer radial and vertical boundaries unnecessary.) 

The base resolution of our simulations amounts to ten grid cells per astronomical unit. On top of that, both static and adaptive mesh refinement were employed. We allowed for up to six levels of refinement, with every level corresponding to a doubling of the resolution. Thus, the maximum resolution is equal to~$320$ cells per astronomical unit. In addition to simulations with these fiducial base and maximum resolutions, we conducted two simulations in which both are doubled to investigate the resolution dependence of our findings. Adaptive mesh refinement was applied to blocks of~$10\times 10$ cells when the number of dust particles in any cell inside these blocks exceeded ten. Blocks were derefined, on the other hand, if no particles were left in a cell. This allows us to resolve dust concentration owing to the streaming instability and the vertical shear instability in the mid-plane layer while minimising the impact of the layers away from the mid-plane that are void of dust on the computational cost of the simulations.

\begin{figure}[t]
  \centering
  \includegraphics[width=\columnwidth]{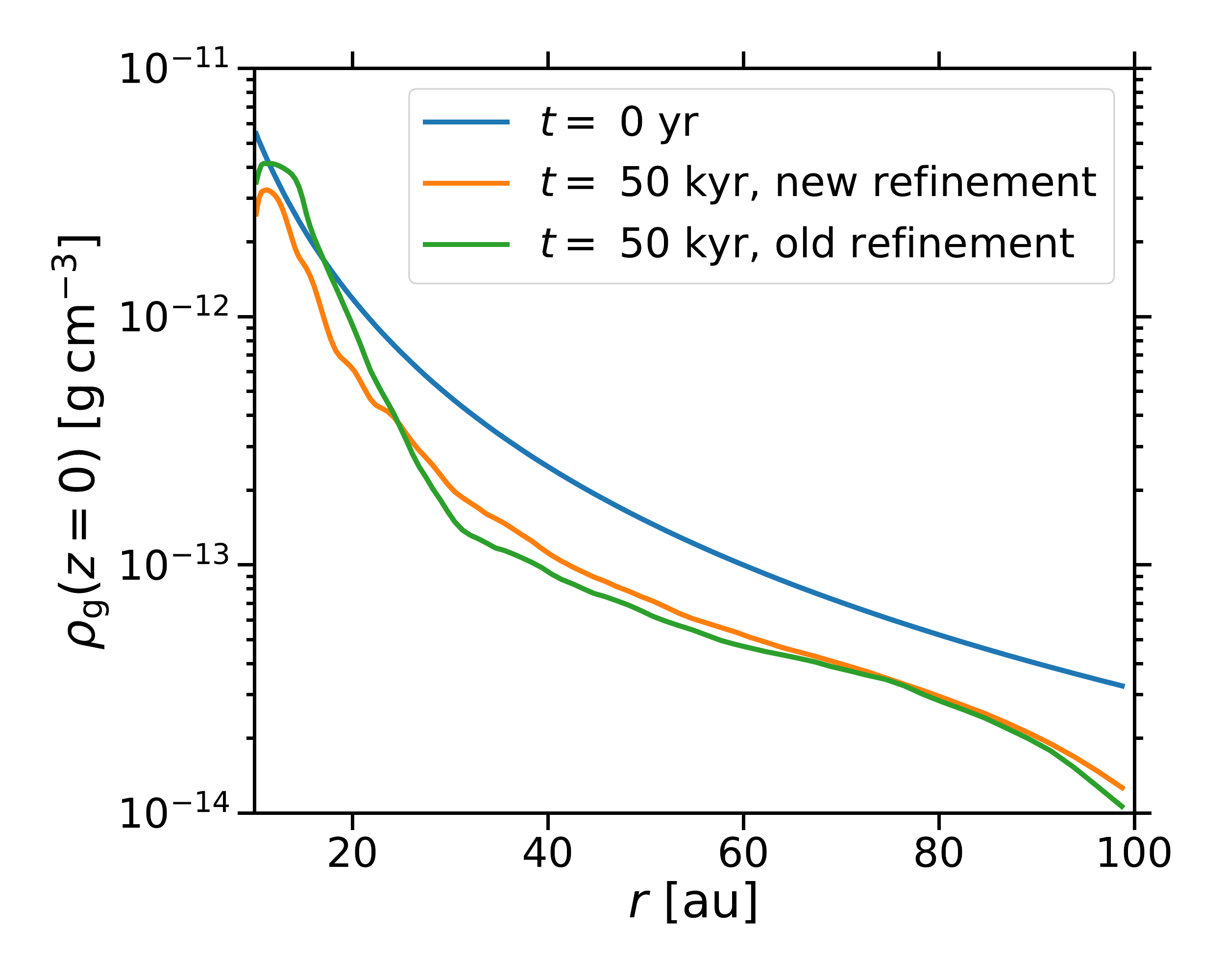}
  \caption{Gas density~$\rho_{\gas}$ in the mid-plane (at a height~\mbox{$z=0$}) as a function of the radius~$r$. The blue line shows the initial density profile, while the orange and green line depict the profiles after~$50~\kyr$ in simulations including the static mesh refinement employed in \citetalias{Schafer2020} and in this study, respectively. Notably, a considerable bump at a radius of~${\sim}30~\au$ is present in the former profile, but not in the latter one.}
  \label{fig:mid-plane_gas_density}
\end{figure}

The usage of static mesh refinement constitutes the only difference between the numerical models we used in \citetalias{Schafer2020} and in this study. In \citetalias{Schafer2020}, the resolution was increased by one or two refinement levels, respectively, where the gas density exceeds~$1\%$ or~$10\%$ of the mid-plane density at the inner radial domain boundary. Here, we instead enhanced the resolution by one or two refinement levels, respectively, within~$5~\au$ and~$1~\au$ above and below the mid-plane. Figure~\ref{fig:mid-plane_gas_density} shows that this modification of the static refinement prevents the formation of a significant bump in the gas density at a radius of~${\sim}30~\au$ before the dust is introduced in simulations of the vertical shear instability only and of the scenario \textit{SIafterVSI}. As is evident from Fig.~8 of \citetalias{Schafer2020}, the presence of this bump entails a large gap in the dust density. This gap does not form in the simulations we present in this paper (see Fig.~\ref{fig:surface_density_azimuthal_velocity}).

\subsection{Gas}
The initial gas density decreases both with the radial distance~$r$ to the star,
\begin{equation}
\rho_{\rm g}(z=0)=5.62\times10^{-12}~\g\,\cm^{-3}~\left(\frac{r}{10~\au}\right)^{-9/4},
\end{equation}
and with the height~$z$ above or below the mid-plane,
\begin{equation}
\rho_{\rm g}=\rho_{\rm g}(z=0)\exp\left[-\frac{\gamma GM_{\rm S}}{c_{\rm s}^2}\left(\frac{1}{r}-\frac{1}{\sqrt{r^2+z^2}}\right)\right],
\end{equation}
where~$G$ is the gravitational constant,~\mbox{$M_{\rm S}=1~M_{\odot}$} the stellar mass,~\mbox{$c_{\rm s}=(\gamma RT/\mu)^{1/2}$} the sound speed,~$\gamma$ the adiabatic index,~$R$ the ideal gas constant,~$T$ the temperature, and~\mbox{$\mu=2.33$} the mean molecular weight. We use the subscripts~$\gas$ and~$\dust$ to refer to gas and dust, respectively. The radial density gradient is shallower than that in the minimum mass solar nebula model, though the mid-plane density at~\mbox{$r=10~\au$} is about two times higher \citep{Hayashi1981}. The vertical density gradient ensures hydrostatic equilibrium in this dimension. We define the surface density as the integral of the density from~$-1$ gas scale height to~$1$ gas scale height (rather than from~$-\infty$ to~$\infty$). It is thus initially given by
\begin{equation}
\Sigma_{\rm g}=\int_{-1~H_{\rm g}}^{1~H_{\rm g}}\rho_{\rm g}~{\rm d}z=10^2~\g\,\cm^{-2}~\left(\frac{r}{10~\au}\right)^{-1},
\end{equation}
where the gas scale height
\begin{equation}
H_{\rm g}=\sqrt{\frac{c_{\rm s}^2r^3(2\gamma GM_{\rm S}-c_{\rm s}^2r)}{(c_{\rm s}^2r-\gamma GM_{\rm S})^2}}=0.846~\au~\left(\frac{r}{10~\au}\right)^{5/4}.
\end{equation}

We adopt the initial gas temperature from the minimum mass solar nebula model \citep{Hayashi1981},
\begin{equation}
T=88.5~\K~\left(\frac{r}{10~\au}\right)^{-1/2}.
\label{eq:temperature}
\end{equation}
This radial temperature gradient causes a height-dependence of the orbital velocity and consequently gives rise to the vertical shear instability. In all simulations of this instability (either in combination with the streaming instability or in isolation), an isothermal equation of state was employed,
\begin{equation}
P=\frac{RT}{\mu}\rho_{\rm g},
\end{equation}
because an infinitely short gas cooling timescale provides ideal conditions for the instability \citep{Nelson2013, Lin2015}. Here,~$P$ is the pressure. On the other hand, in simulations of the streaming instability only we used an adiabatic equation of state,
\begin{equation}
P=K\rho_{\rm g}^{\gamma},
\end{equation}
where~\mbox{$K=RT\rho_{\gas}^{1-\gamma}/\mu$} is the polytropic constant -- constant in time, but varying in space with the temperature and density distributions -- and the adiabatic index~\mbox{$\gamma=5/3$}. This is since under these conditions the vertical shear instability is quenched by vertical buoyancy.

The radial gradients in gas density and temperature entail a pressure gradient, which is necessary for the streaming instability to be active. We express the strength of the pressure gradient in terms of the dimensionless parameter \citep{Bai2010b}
\begin{equation}
\Pi=-\frac{1}{2c_{\rm s}\rho_{\rm g}\Omega_{\rm K}}\frac{{\rm d}P}{{\rm d}r},
\label{eq:pressure_gradient}
\end{equation}
where~$\Omega_{\rm K}$ is the Keplerian orbital frequency. In the mid-plane, this parameter amounts to
\begin{equation}
\Pi(z=0)=8.2\times10^{-2}\left(\frac{r}{10~\au}\right)^{1/4}.
\label{eq:pressure_gradient_parameter}
\end{equation}
The initial orbital velocity is chosen such that this pressure gradient and the centrifugal force balance the radial stellar gravity.

\subsection{Dust}
As is common in numerical models of the streaming instability including Lagrangian particles to represent the dust \citep{Youdin2007a, Bai2010a}, every particle in our simulations possesses the total mass and momentum of a huge number of the dust aggregates that are present in protoplanetary disks, but the drag coupling to the gas of a single such aggregate. Our simulations include~$10^6$ dust particles. Initially, these are uniformly distributed in the radial dimension while their vertical positions are randomly sampled from a Gaussian distribution with a scale height equal to~$10\%$ of the gas scale height. The noise in the vertical distribution seeds the streaming instability. We find our results to be independent of whether~$5\times10^5$ or~$10^6$ particles are simulated as well as of the random seed, and discuss the dependence on the initial scale height in \citetalias{Schafer2020}.

Because of their homogeneous radial distribution, the particle mass is given by
\begin{equation}
m_{\dust}=\frac{1}{N_{\dust}}\int_{L_r} 2\pi r\Sigma_{\dust}~{\rm d}r,
\end{equation}
where~{$N_{\dust}=10^6$} is the number of particles and~\mbox{$L_r=90~\au$} the radial domain extent. Since we choose the dust-to-gas surface density ratio to be constant initially and the gas surface density is inversely proportional to the radius, the mass of every particle thus amounts to
\begin{equation}
m_{\rm d}=1.27\times10^{24}~\g~\left(\frac{\Sigma_{\dust}/\Sigma_{\gas}}{0.01}\right).
\end{equation}
We consider surface density ratios of~$0.5\%$, $1\%$, and~$2\%$. These ratios are equal to the mean dust-to-gas mass ratio in the Milky Way interstellar medium or smaller or greater by a factor of two. Nonetheless, they are low when compared to the observed mass ratios in protoplanetary disks (see Sect.~\ref{sect:introduction}).

\begin{figure}[t]
  \centering
  \includegraphics[width=\columnwidth]{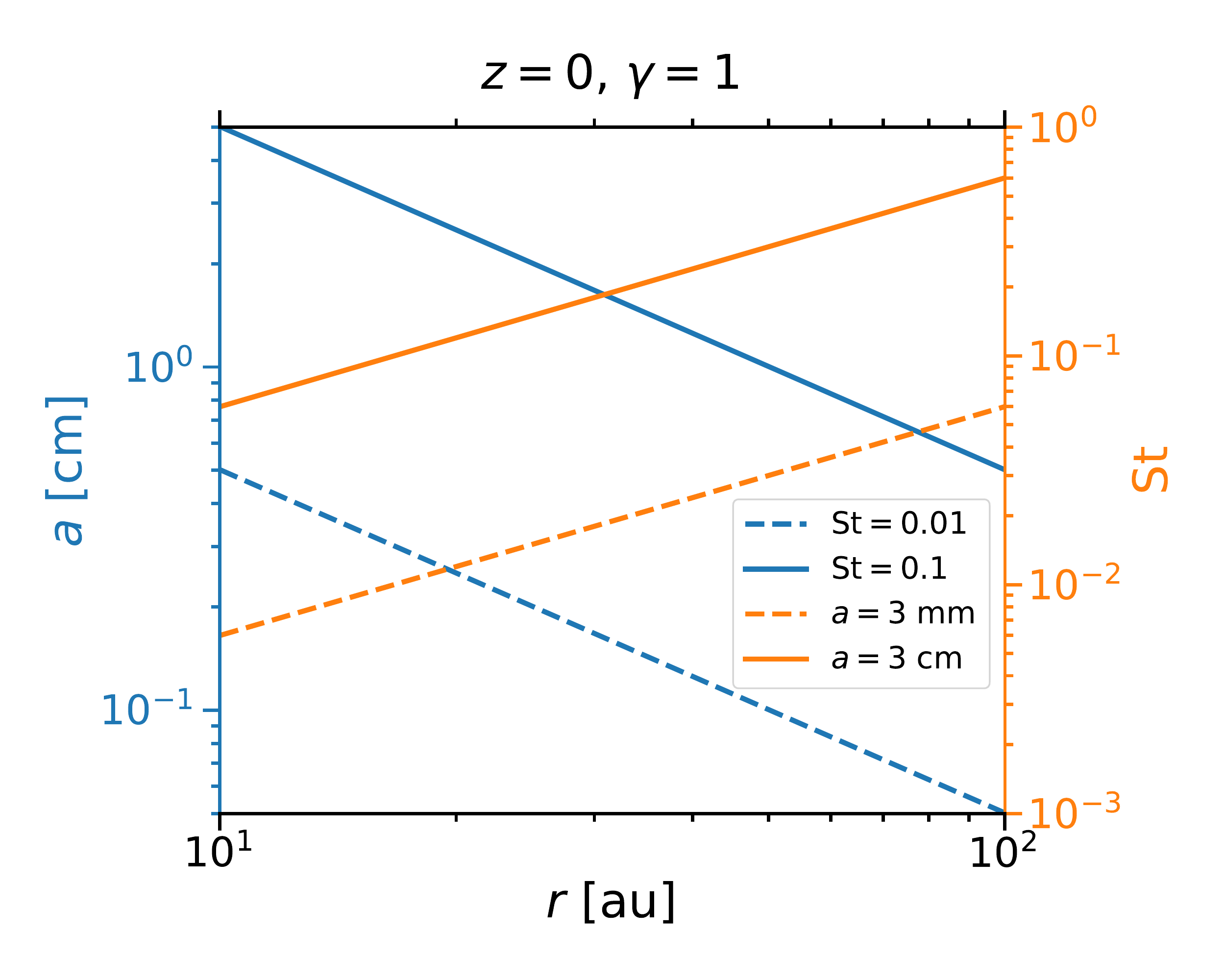}
  \caption{Dust size~$a$ (left ordinate) and Stokes number~$\St$ (right ordinate) as functions of the radius~$r$. Conversion relations between dust size and Stokes number in the mid-plane of our protoplanetary disk model (Eqs.~\ref{eq:Stokes_number} and~\ref{eq:dust_size}) are shown for the two dust sizes (orange lines) and the two Stokes numbers (blue lines) that we simulated, assuming an isothermal equation of state.}
  \label{fig:dust_size_Stokes_number}
\end{figure}

All particles in a given simulation are of the same size or Stokes number, respectively, with sizes of~$3~\mm$ and~$3~\cm$ as well as Stokes numbers of~$0.01$ and~$0.1$ being taken into account. As detailed in Sect.~\ref{sect:introduction}, these are consistent with the sizes inferred from the opacity spectral index of the dust emission from protoplanetary disks, but larger than the ones derived from the polarisation of the emission. When considering the disk mid-plane, we can convert from dust size to Stokes number and vice versa using the relations 
\begin{align}
\St(z=0)&=t_{\rm d,stop}(z=0)\Omega_{\rm K}(z=0)=\frac{a\rho_{\rm s}}{c_{\rm s}\rho_{\rm g}(z=0)}\Omega_{\rm K}(z=0)\nonumber\\
&=6\times10^{-3}~\frac{1}{\sqrt{\gamma}}\left(\frac{a}{3~\mm}\right)\left(\frac{r}{10~\au}\right)\text{ and}
\label{eq:Stokes_number}
\end{align}
\begin{equation}
a(z=0)=5.02~\mm~\sqrt{\gamma}\left(\frac{\St}{0.01}\right)\left(\frac{r}{10~\au}\right)^{-1},
\label{eq:dust_size}
\end{equation}
where~$t_{\rm d,stop}$ is the dust stopping time,~$a$ the dust size, and~\mbox{$\rho_{\rm s}=1~\g\,\cm^{-3}$} is the dust material density. Here, we utilise that the dust size is smaller than the gas mean free path length at all gas densities in our model and therefore calculate the dust stopping time in the Epstein regime. Figure~\ref{fig:dust_size_Stokes_number} illustrates these relations exemplarily for our model with an isothermal equation of state. The particles initially orbit with the Keplerian speed.

\section{Dust concentration}
\label{sect:dust_concentration}
\subsection{Synopsis of relevant results from \citet{Schafer2020}}
As in \citetalias{Schafer2020}, we investigate three scenarios: one in which the streaming instability operates in isolation and two in which both this instability and the vertical shear instability are active. In \emph{SIwhileVSI}, both instabilities develop simultaneously. This scenario is similar to that including only the streaming instability in that the turbulence in the mid-plane dust layer is predominantly driven by the streaming instability, with the dust scale height amounting to~${\sim}1\%$ of the gas scale height.

In \emph{SIafterVSI}, on the other hand, the vertical shear instability has saturated before the streaming instability begins to grow. The vertical shear instability remains the primary source of turbulence in the dust layer in this scenario, though the streaming instability causes turbulence locally in dust overdensities. Since the vertical shear instability causes stronger turbulent motions in the vertical dimension than the streaming instability, the dust scale height is equal to~${\sim}10\%$ of the gas scale height in this scenario.

Nonetheless, in \citetalias{Schafer2020} we show that dust overdensities are more prominent and the maximum dust-to-gas volume density ratio is higher in the scenario \emph{SIafterVSI} than in the scenario \emph{SIwhileVSI}. We conjecture that dust overdensities caused by the vertical shear instability trigger the streaming instability and are reinforced by it. In this section, we investigate dust concentration in the three scenarios in more detail.

\subsection{Spatial dust distribution}
\label{sect:dust_distribution}
\begin{figure*}[t]
  \centering
  \includegraphics[width=\textwidth]{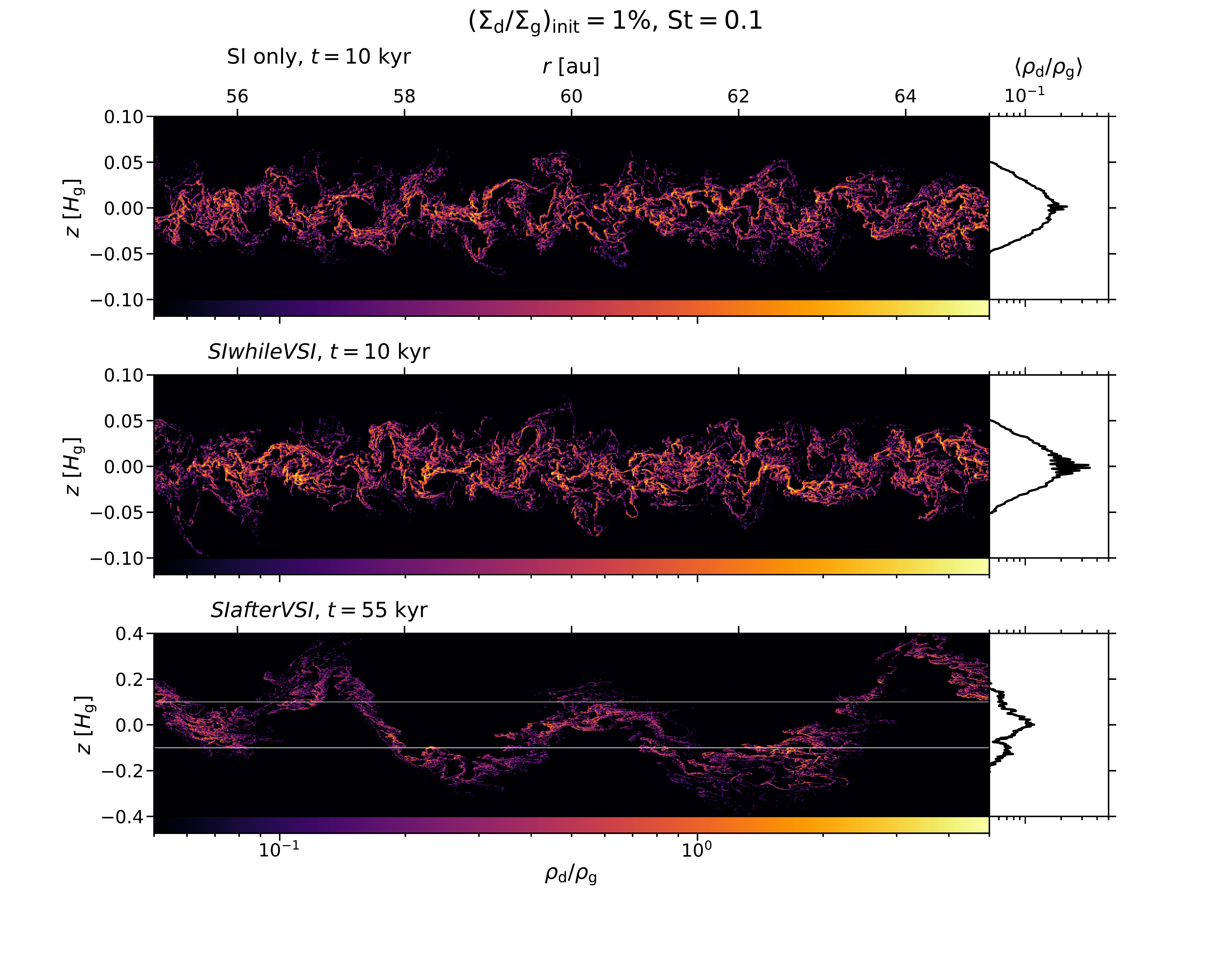}
  \vspace{-1.2cm}
  \caption{Ratio of dust~$\rho_{\dust}$ to gas volume density~$\rho_{\gas}$ as a function of radius~$r$ and height~$z$, the latter given in units of gas scale heights~$H_{\gas}$ (left panels). The average over the radius range shown in the left panels is depicted as a function of height in the right panels. The top, middle, and lower panels show the dust-to-gas volume density ratio at the end of simulations of the streaming instability in isolation, the scenario \emph{SIwhileVSI}, and the scenario \emph{SIafterVSI}, respectively, with an initial dust-to-gas surface density ratio of~$1\%$ and a Stokes number of~$0.1$. The horizontal white lines in the lower panel illustrate the range of heights depicted in the upper and middle panels. In all simulations, local maxima of the volume density ratio are distributed over the entire vertical dimension of the dust layer. While the vertical dust distribution in the simulations of the streaming instability only and of the scenario \emph{SIwhileVSI} is largely Gaussian, the dust layer possesses the form of a wave rather than a Gaussian shape in the \emph{SIafterVSI} simulation.}
  \label{fig:density_ratio}
\end{figure*}

To begin with, we explore the vertical distribution of the dust in our models. Figure~\ref{fig:density_ratio} depicts the dust-to-gas volume density ratio in simulations of the streaming instability only, of the scenario \emph{SIwhileVSI}, and of the scenario \emph{SIafterVSI} with the same initial dust-to-gas surface density ratio of~$1\%$ and Stokes number of~$0.1$. In all three simulations, local maxima of the volume density ratio are not concentrated in the mid-plane, but present at all heights inside the dust layer. Nevertheless, in the simulations of the streaming instability alone and of the scenario \emph{SIwhileVSI} the dust layer resembles a Gaussian distribution, with the radial average of the volume density ratio reaching its maximum near or in the mid-plane.

In contrast, in the \emph{SIafterVSI} simulation the shape of the layer is wave-like and markedly deviates from a Gaussian. While the dust scale height is considerably greater in this scenario than in the other two, the vertical thickness of the dust layer is not. This illustrates that stronger large-scale turbulence regulates the scale height while weaker small-scale turbulence drives the internal diffusion in the layer. The millimetre-sized dust in the three-dimensional models of the vertical shear instability that are presented by \citet{Flock2017} and \citet{Flock2020} possesses a similar wave-shaped distribution, both in the radial-vertical plane as in our two-dimensional model and in the azimuthal-vertical plane (see Figs.~8 and~9 of \citealt{Flock2017} and Figs.~7 and~D1 of \citealt{Flock2020}.)

\begin{figure*}[t]
  \centering
  \includegraphics[width=\textwidth]{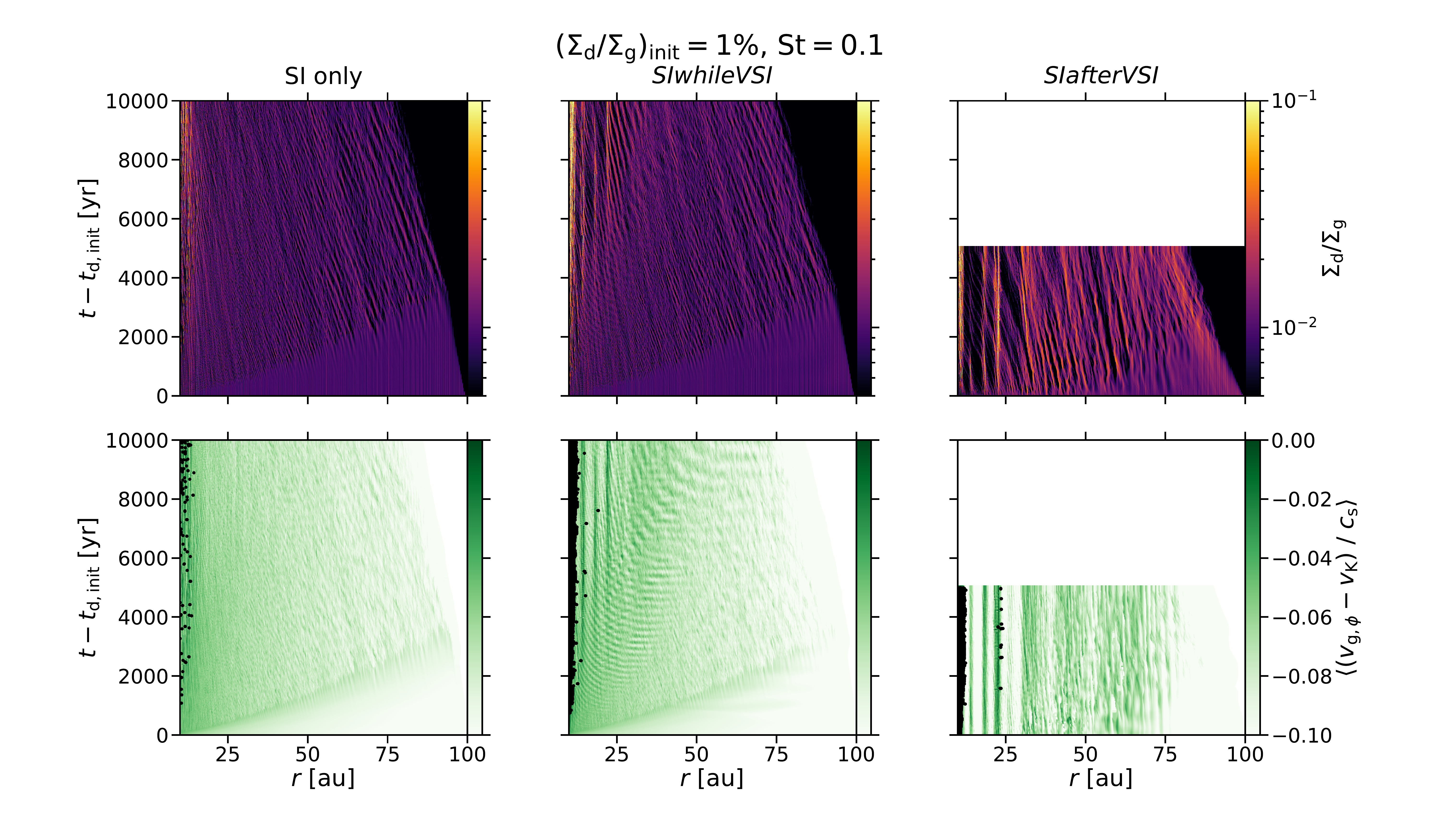}
  \caption{Ratio of dust~$\Sigma_{\dust}$ to gas surface density~$\Sigma_{\gas}$ (upper panels) and difference between azimuthal gas velocity~$v_{\gas,\phi}$ and Keplerian velocity~$v_{\rm K}$ (lower panels). Both quantities are depicted as a function of radius~$r$ and time~$t$ after the initialisation of the dust particles at~$t_{\textrm{d,init}}$. The velocity difference is averaged over the vertical domain size, weighted by the dust mass, and expressed as a Mach number. In the absence of turbulence, this quantity is equal to the negative of the parameter~$\Pi$ as defined in Eq.~\ref{eq:pressure_gradient}.  Black dots indicate that the azimuthal velocity is equal to or greater than the Keplerian velocity. Enhancements in the dust-to-gas surface density ratio and in the difference between azimuthal gas velocity and Keplerian velocity are larger at late times~\mbox{($t-t_{\textrm{d,init}}>5~\kyr$)} and small radii~\mbox{($r<40~\au$)} in \emph{SIwhileVSI} than in the simulation of the streaming instability in isolation, and greatest at all times and radii in \emph{SIafterVSI}. Surface density ratio and velocity difference can be seen to generally be correlated.}
  \label{fig:surface_density_azimuthal_velocity}
\end{figure*}

As shown also in \citetalias{Schafer2020}, we find that dust overdensities are greater in the scenario \emph{SIafterVSI} than in the scenario \emph{SIwhileVSI}. This is illustrated in the upper panels of Fig.~\ref{fig:surface_density_azimuthal_velocity}, which depict the dust-to-gas surface density ratio in the same simulations as shown in Fig.~\ref{fig:density_ratio}. From the figure, it can further be seen that after~$5~\kyr$ and at radii less than~$40~\au$ dust overdensities are larger in \emph{SIwhileVSI} than in the simulation of the streaming instability only. This can be explained by the vertical shear instability growing over several thousand years -- it takes about 30 orbital periods to saturate \citep{Stoll2014, Flock2017}, longer than the streaming instability \citepalias{Schafer2020} -- until it starts to influence the dust mid-plane layer and enhance dust overdensities in it.

\begin{figure*}[t]
  \centering
  \includegraphics[width=\textwidth]{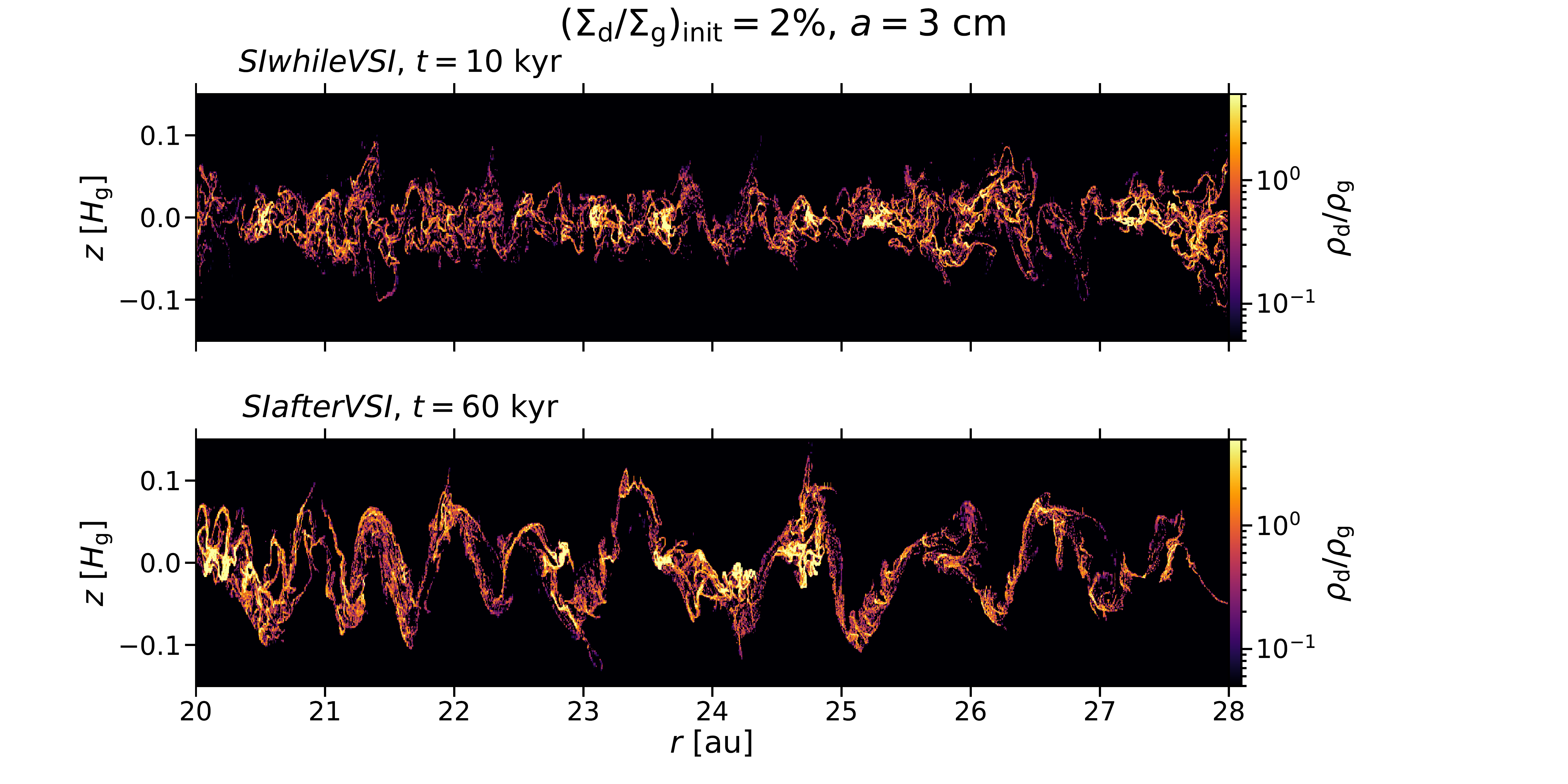}
  \caption{Dust-to-gas volume density ratio~$\rho_{\dust}/\rho_{\gas}$ as a function of radius~$r$ and height~$z$ in simulations of the scenarios \emph{SIwhileVSI} (top panel) and \emph{SIafterVSI} (bottom panel) with an initial dust-to-gas surface density ratio of~$2\%$ and a dust size of~$3~\cm$. The volume density ratio is depicted at the end of the simulations~$10~\kyr$ after the initialisation of the dust particles. The vertical dust distribution in the \emph{SIwhileVSI} simulation resembles a Gaussian, but is wave-like in the \emph{SIafterVSI} simulation.}
  \label{fig:density_ratio_SIwhileVSI_SIafterVSI}
\end{figure*}

The question arises whether comparable simulations of the scenarios \emph{SIwhileVSI} and \emph{SIafterVSI} would eventually evolve into the same state if they were continued sufficiently long. To address this question, we show the dust-to-gas volume density ratio in simulations of these two scenarios with an initial dust-to-gas surface density ratio of~$2\%$ and a dust size of~$3~\cm$ in Fig.~\ref{fig:density_ratio_SIwhileVSI_SIafterVSI}. In the figure, the dust distribution is depicted~$10~\kyr$ or~${\sim}100$ local orbits after the dust is introduced into the simulations (at the beginning in the case of the \emph{SIwhileVSI} simulation and after~$50~\kyr$ in the case of the \emph{SIafterVSI} simulation). As can be seen, the dust layer in \emph{SIwhileVSI} continues to possess a Gaussian shape, while the dust layer in \emph{SIafterVSI} still does not. We can not exclude that the two simulations would develop into a similar state on even longer timescales, though.

In summary, the vertical shear instability and the streaming instability together induce stronger dust concentration than the streaming instability alone, independent of whether the vertical shear instability (scenario \emph{SIafterVSI}) or the streaming instability (scenario \emph{SIwhileVSI}) is the dominant source of turbulence in the dust layer. The concentration is strongest in the scenario \emph{SIafterVSI}, despite the vertical shear instability inducing a larger dust scale height in this scenario than the streaming instability in the other two scenarios as this larger scale height is not synonymous with a greater vertical thickness of the dust layer.

\subsection{Connection between dust concentration and pressure bumps}
The lower panels of Figure~\ref{fig:surface_density_azimuthal_velocity} depict the difference between the azimuthal gas velocity and the Keplerian velocity. A correlation of this velocity difference and the dust-to-gas surface density ratio, which is shown in the upper panels, is apparent in our model of the streaming instability in isolation (left panels). This can be explained by the drag exerted by the dust on the gas causing the azimuthal gas velocity to be closer to the Keplerian velocity where the dust-to-gas volume density ratio is enhanced.

Nevertheless, an increased azimuthal gas velocity is not only an indicator of dust overdensities, but also of pressure bumps, that is local deviations from the global gas pressure gradient. This is since the velocity depends linearly on the pressure gradient and is equal to the Keplerian velocity if the pressure attains a local maximum and the gradient vanishes (assuming that turbulence is negligible). We note that \citet{Li2018} show that the streaming instability causes weak pressure bumps, but independently of dust concentration. Moreover, \citet{Yang2014} find a weak anti-correlation between dust and gas density in their streaming instability simulations, while dust overdensities and local increases in the azimuthal gas velocity are correlated in our simulations.

On the other hand, previous studies show that dust accumulates in vortices \citep{Flock2017, Flock2020, Lehmann2022} and short-lived pressure fluctuations \citep{Stoll2016} induced by the vertical shear instability. While vortices can not form in our two-dimensional model, it can already be seen from Fig.~\ref{fig:mid-plane_gas_density} that the vertical shear instability indeed gives rise to pressure bumps. A comparison of the left and middle panels of Fig.~\ref{fig:surface_density_azimuthal_velocity} reveals that thus both local enhancements in the azimuthal gas velocity and dust overdensities are more pronounced at late times and small radii -- when the vertical shear instability has grown sufficiently to influence the dynamics in the dust mid-plane layer -- in the scenario \emph{SIwhileVSI} than if only the streaming instability is simulated.

\begin{figure*}[t]
  \centering
  \includegraphics[width=0.8\textwidth]{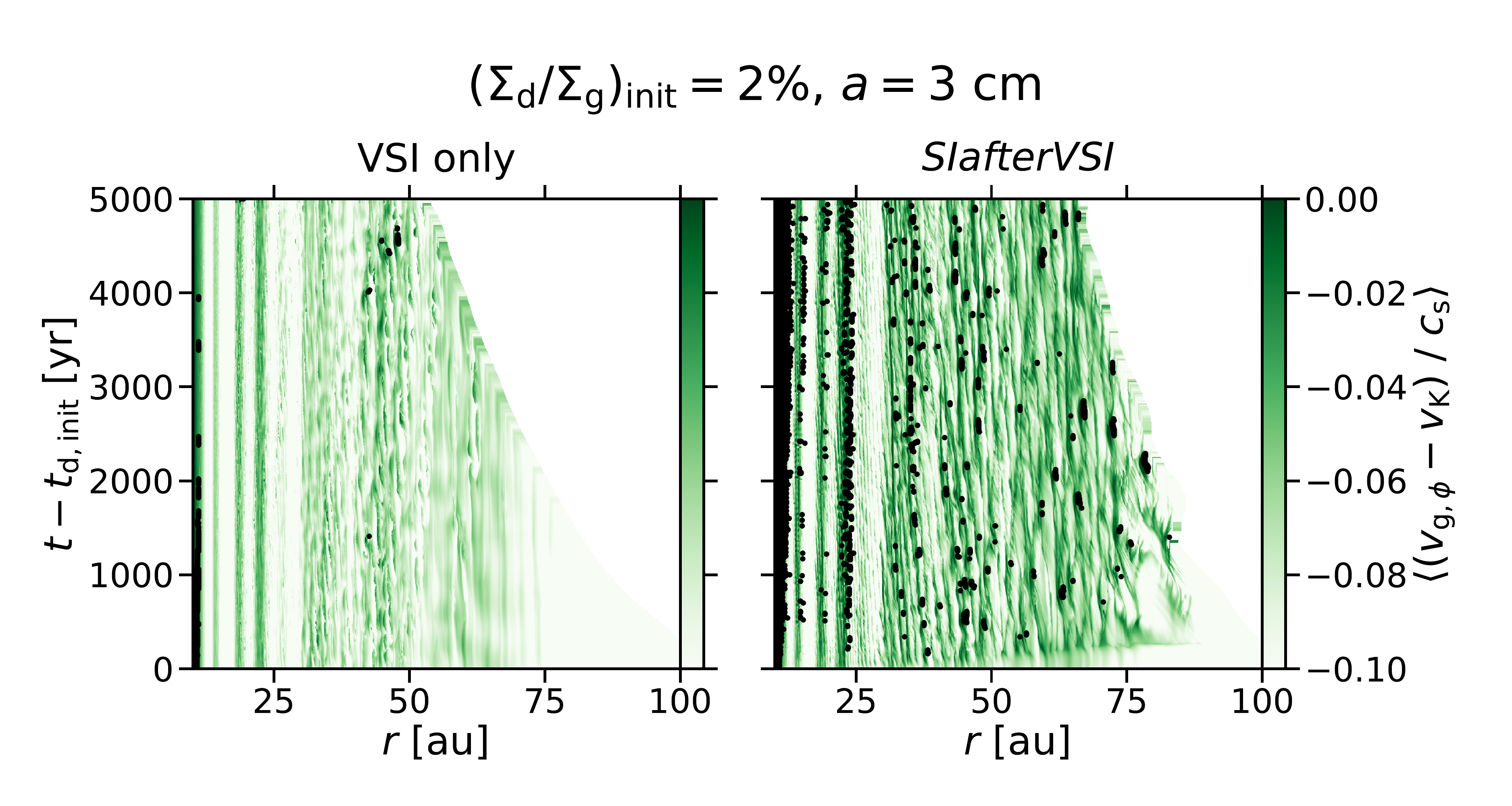}
  \caption{Dust-mass-weighted mean deviation of azimuthal gas velocity from Keplerian velocity~$v_{\gas,\phi}-v_{\rm K}$, normalised by the sound speed~$c_{\textrm{s}}$, as a function of radius~$r$ and time after the dust particle initialisation~$t-t_{\textrm{d,init}}$. The left and right panels depict simulations of the vertical shear instability only and of the scenario \emph{SIafterVSI}, respectively. Local increases in the azimuthal gas velocity are stronger, with the velocity more frequently exceeding the Keplerian velocity, and especially at~\mbox{$r>30~\au$} also more persistent in \emph{SIafterVSI}.}
  \label{fig:azimuthal_velocity}
\end{figure*}

Like the dust-to-gas surface density, the azimuthal gas velocity is most strongly locally enhanced in the scenario \emph{SIafterVSI} (right panels) as compared with the other two scenarios. While in these scenarios turbulence in the dust layer is predominantly driven by the streaming instability, it is mainly caused by the vertical shear instability in \emph{SIafterVSI}. Nonetheless, augmentations of the azimuthal gas velocity are greater and more long-lived in this scenario than the augmentations that result from the transient pressure bumps caused by the vertical shear instabiltity. This can be seen from Fig.~\ref{fig:azimuthal_velocity}, which shows the deviation of the azimuthal gas velocity from the Keplerian velocity in our simulation of only the vertical shear instability and the simulation of the scenario SIafterVSI with the same initial dust-to-gas surface density ratio and dust size.

That is, local enhancements in the azimuthal gas velocity are stronger in the scenario \emph{SIafterVSI} than if either only the streaming instability or only the vertical shear instability are considered. In line with what we speculate in \citetalias{Schafer2020}, we conclude that in \emph{SIafterVSI} dust accumulates in pressure bumps caused by the vertical shear instability, with these accumulations acting as seeds for and being reinforced by the streaming instability. Indeed, in \emph{SIafterVSI} we show that while turbulence in the dust layer is primarily driven by the vertical shear instability in this scenario, it is induced by the streaming instability locally in dust overdensities.

We do not further discuss our model of only the vertical shear instability, and in particular dust concentration in it, in this paper. This is because simulating only this instability requires neglecting the drag exerted by the dust on the gas since otherwise the streaming instability, too, would be active. Excluding this drag is only justified if the dust density is much less than the gas density. However, this condition can not be reconciled with our finding that dust concentration is high when the drag of the dust onto the gas is taken into account.

\section{Metrics to establish planetesimal formation}
\label{sect:metrics}
The formation of planetesimals can be observed in three-dimensional simulations of the streaming instability including dust self-gravity \citep[e.g.][]{Johansen2007b, Johansen2009, Johansen2011}. However, because the computational cost of such three-dimensional simulations is prohibitive, two-dimensional models without self-gravity are employed to explore which combinations of dust-to-gas surface density ratio and dust size or Stokes number provide the necessary conditions for the streaming instability to induce planetesimal formation \citep{Carrera2015, Yang2017, Li2021}. 

These parameter studies -- including the one presented in this paper -- aim to establish whether dust concentration owing to the streaming instability is sufficiently strong that it would lead to planetesimal formation in equivalent three-dimensional models with self-gravity. \citet{Carrera2015} assume that this is the case if the time-averaged dust surface density distribution deviates sufficiently from a uniform distribution. \citet{Yang2017}, on the other hand, investigate whether strong dust clumping can be seen from the spatial distribution of the dust-to-gas volume density ratio. Finally, \citet{Li2021} examine whether the maximum dust density exceeds the Roche density.

However, an important caveat to any approach involving the maximum dust-to-gas volume density ratio is that this quantity is inherently stochastic. A measured high density ratio might be a numerical artifact, attained in very few or only a single grid cell. And even if it is physical, it can only be used to establish if planetesimals would potentially form, but does not reveal any information about the properties of the planetesimals. Moreover, when comparing to the Roche density one needs to bear in mind that overdensities which barely exceed the Roche density might only be transient if they are not sufficiently gravitational unstable to overcome turbulent diffusion.\footnote{\citet{Gerbig2020} and \citet{Klahr2020} propose criteria for collapse that involve both the overdensities exceeding the Roche density and their self-gravity overcoming diffusion.} We address these caveats below.

\subsection{Correlation between maximum and mean dust-to-gas density ratio and strong clumping}
\label{sect:density_ratio}
\begin{figure*}[t]
  \centering
  \includegraphics[width=\textwidth]{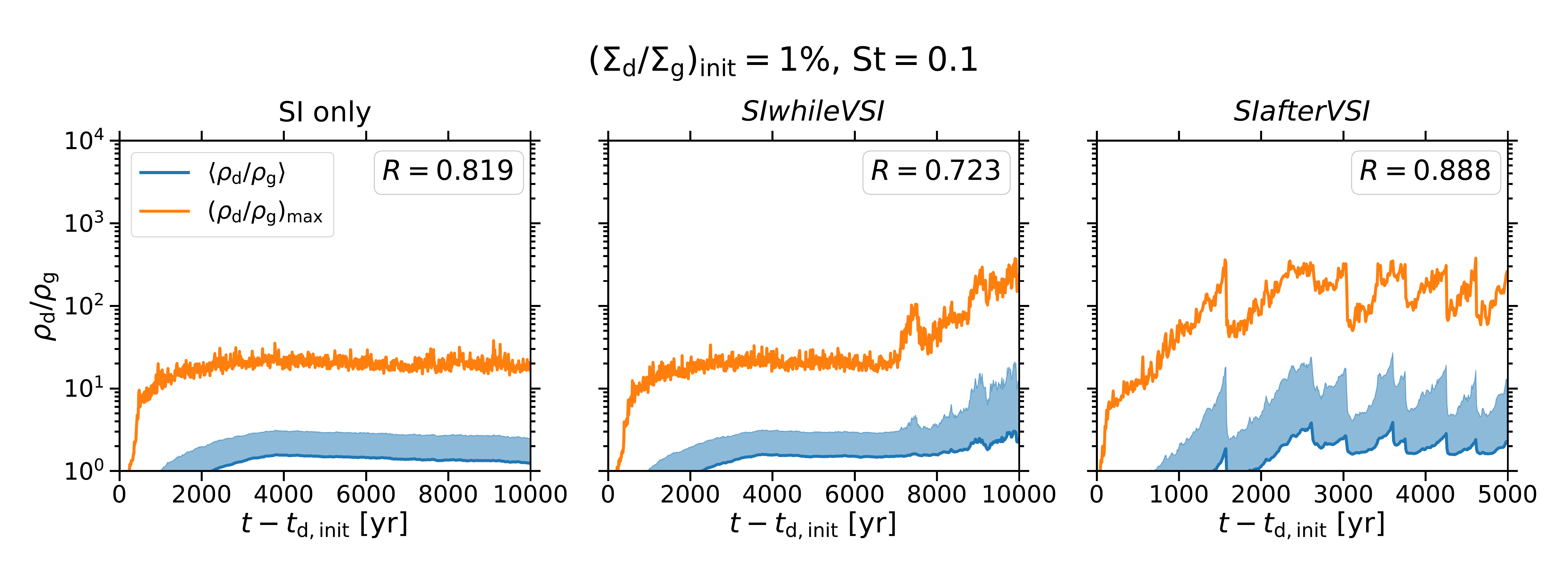}
  \caption{Maximum (orange lines) as well as dust-mass-weighted average (blue lines) and standard deviation (blue-shaded areas) of dust-to-gas volume density ratio~$\rho_{\dust}/\rho_{\gas}$ as a function of time after the dust particle initialisation~$t-t_{\textrm{d,init}}$. The maximum, mean, and standard deviation are well-correlated, with the Pearson correlation coefficients\textsuperscript{\ref{footnote}} of maximum and mean being given in the top right of the panels. Three phases in the evolution of the volume density ratio in the \emph{SIwhileVSI} simulation can be distinguished: an initial phase of sedimentation, followed by a quasi-steady phase, and a phase of strong clumping characterised by strong variations. In comparison, the simulation of the streaming instability never evolves into the strong-clumping phase, while the \emph{SIafterVSI} simulation skips the quasi-steady phase.}
  \label{fig:density_ratio_maximum_mean} 
\end{figure*}

Figure~\ref{fig:density_ratio_maximum_mean} depicts the evolution of the maximum, mean, and standard deviation of the dust-to-gas volume density ratio, where the mean is weighted by the dust mass, in the same simulations as shown in Figs.~\ref{fig:density_ratio} and~\ref{fig:surface_density_azimuthal_velocity}. Similar to in the simulations of the streaming instability presented by \citet{Yang2017} and \citet{Li2021}, the evolution of the maximum volume density ratio can be divided into three phases: dust settling and formation of an equilibrium mid-plane layer, a subsequent quasi-steady state, and finally a phase of strong clumping. The latter is characterised by greater maxima and large variations that reflect the formation and dissolution of prominent overdensities, as is evident from comparing this figure with the upper panels of Fig.~\ref{fig:surface_density_azimuthal_velocity}.

It is this strong-clumping phase that is associated with the formation of planetesimals \citep{Johansen2015, Yang2017, Li2021}. Only the simulations of \emph{SIwhileVSI} and \emph{SIafterVSI} evolve into such a phase, however, while the simulation of the streaming instability alone remains in an equilibrium state. Indeed, as we show in Sect.~\ref{sect:dust_distribution}, dust overdensities caused by the vertical shear instability and the streaming instability in combination are larger than the ones induced by the streaming instability in isolation. While in the \emph{SIafterVSI} simulation sedimentation is directly followed by strong clumping, the \emph{SIwhileVSI} simulation undergoes a quasi-steady phase until strong clumping sets in when the vertical shear instability starts to affect the dust layer.

There is a strong correlation between the maximum of the dust-to-gas volume density ratio and its mean and standard deviation in all models. This shows that the maximum is in fact representative of the dust clumping behaviour not only in a single or a few cells, but globally in our models that cover the mid-plane layer on the scale of entire protoplanetary disks. As can be seen in Fig~\ref{fig:density_ratio_maximum_mean}, after sedimentation the mean and maximum are nearly constant in the simulation of the streaming instability only, with the Pearson correlation coefficient\footnote{\label{footnote}The Pearson correlation coefficient quantifies the linear correlation between two variables and is defined as the covariance of the variables divided by the product of their standard deviations.} of the two amounting to~\mbox{$R=0.819$}. In the \emph{SIafterVSI} simulation, on the other hand, the correlation coefficient is even greater with a value of~\mbox{$R=0.888$} despite greater variations in mean and maximum. This is because the building and breaking up of overdensities is reflected not only in the maximum, but also in the average. We find this to be generally true in our models since the median correlation coefficient is equal to~\mbox{$R=0.823$} for simulations which develop into a strong-clumping phase, compared with~\mbox{$R=0.857$} for simulations which do not advance from a quasi-steady state. \citet{Simon2016} find a similar correlation between the maximum dust density and a measure of the dust density dispersion, the ratio of root mean square to mean density.

\subsection{Parameter study of maximum dust-to-gas density ratio}
\label{sect:maximum_density_ratio}
\begin{figure*}[t]
  \centering
  \includegraphics[width=\textwidth]{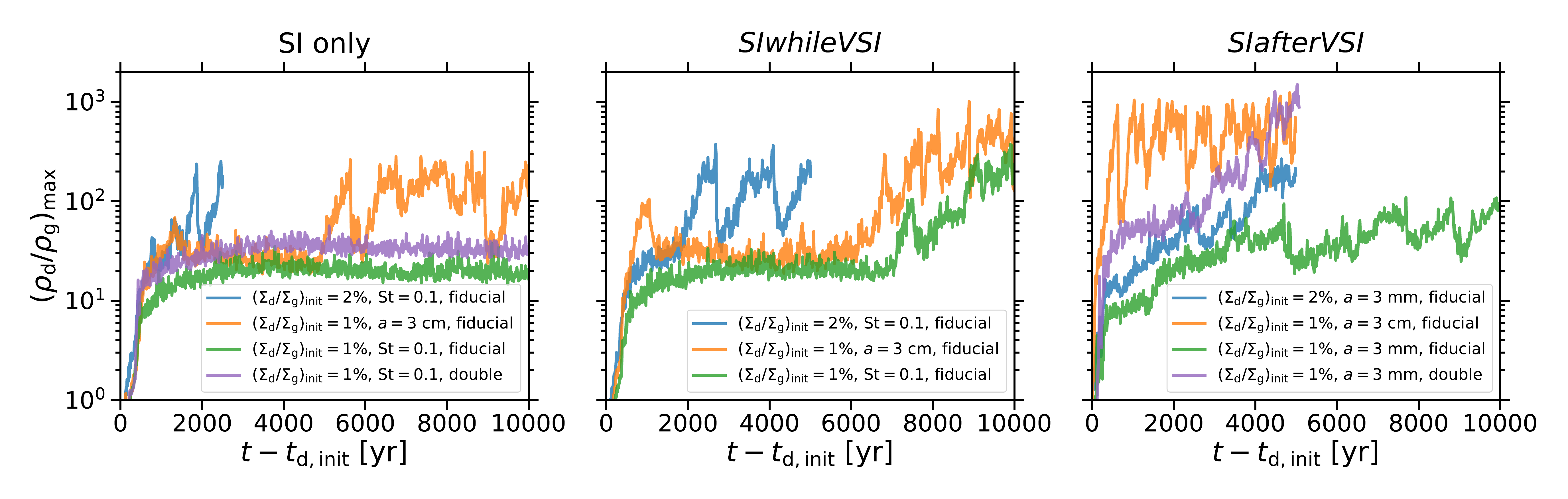}
  \caption{Maximum dust-to-gas volume density ratio~$(\rho_{\dust}/\rho_{\gas})_{\textrm{max}}$ as a function of time after the dust particle initialisation~$t-t_{\textrm{d,init}}$. In the left, middle, and right panels, respectively, simulations of the streaming instability only, of the scenario \emph{SIwhileVSI}, and of the scenario \emph{SIafterVSI} with different initial dust-to-gas surface density ratios~$(\Sigma_{\dust}/\Sigma_{\gas})_{\textrm{init}}$, fixed dust sizes~$a$ or Stokes numbers~$\St$, and resolutions (fiducial or double) can be seen. Increasing any of the three parameters causes the maximum volume density ratio to increase as well. In addition, the maximum  is larger if a dust size of~$a=3~\cm$ rather than a Stokes number of~$0.1$ is considered, as is evident from the left and middle panels. Neither the simulation of the streaming instability with the doubled resolution nor the corresponding one with the fiducial resolution develop into a strong-clumping phase, with the maximum volume density ratio being greater by a factor of a few in the former simulation. In comparison, the maximum is enhanced by more than an order of magnitude during the strong-clumping phase in the double-resolution simulation of \emph{SIafterVSI} than in respective fiducial-resolution simulation.}
  \label{fig:density_ratio_maximum_comparison}
\end{figure*}

\begin{table*}
\caption{Simulation statistics}
\centering
\resizebox{\hsize}{!}{
\begin{tabular}{lcccccc}
\hline
\hline
Scenario&$(\Sigma_{\dust}/\Sigma_{\gas})_{\textrm{init}}$ [\%]\tablefootmark{a}&Dust size\tablefootmark{b}&Base, maximum&Strong&$\langle(\rho_{\dust}/\rho_{\gas})_{\textrm{max}}\rangle$\tablefootmark{c}&$\langle\Delta f_{\dust}(\rho_{\dust}\geq\rho_{\textrm{R}})/\Delta t\rangle$\tablefootmark{d}\\
&&&resolution [$\au^{-1}$]&clumping&[g$\,$cm$^{-3}$]&[$\kyr^{-1}$]\\
\hline
\hline
SI only&0.5&$\St=0.1$&$10$, $320$&\xmark&$10\pm2$&-\\
SI only&0.5&$a=3~\cm$&$10$, $320$&\xmark&$20\pm5$&-\\
SI only&1&$\St=0.1$&$10$, $320$&\xmark&$20\pm3$&-\\
SI only&1&$\St=0.1$&$20$, $640$&\xmark&$34\pm5$&-\\
SI only&1&$a=3~\cm$&$10$, $320$&\cmark&$119\pm59$&$0.057\pm0.046$\\
SI only&2&$a=3~\mm$&$10$, $320$&\xmark&$25\pm3$&-\\
SI only&2&$\St=0.1$&$10$, $320$&\cmark&$78\pm52$&$0.085\pm0.041$\\
SI only&2&$a=3~\cm$&$10$, $320$&\cmark&$126\pm53$&$0.264\pm0.093$\\
\hline
\textit{SIwhileVSI}&0.5&$\St=0.1$&$10$, $320$&\xmark&$13\pm3$&-\\
\textit{SIwhileVSI}&0.5&$a=3~\cm$&$10$, $320$&\xmark&$23\pm6$&-\\
\textit{SIwhileVSI}&1&$\St=0.1$&$10$, $320$&\cmark&$119\pm76$&$0.107\pm0.107$\\
\textit{SIwhileVSI}&1&$a=3~\cm$&$10$, $320$&\cmark&$286\pm169$&$0.174\pm0.090$\\
\textit{SIwhileVSI}&2&$a=3~\mm$&$10$, $320$&\xmark&$30\pm4$&-\\
\textit{SIwhileVSI}&2&$\St=0.1$&$10$, $320$&\cmark&$130\pm66$&$0.118\pm0.046$\\
\textit{SIwhileVSI}&2&$a=3~\cm$&$10$, $320$&\cmark&$1089\pm861$&$0.084\pm0.042$\\
\hline
\textit{SIafterVSI}&0.5&$\St=0.01$&$10$, $320$&\xmark&$0.6\pm0.2$&-\\
\textit{SIafterVSI}&0.5&$a=3~\mm$&$10$, $320$&\xmark&$18\pm4$&-\\
\textit{SIafterVSI}&0.5&$\St=0.1$&$10$, $320$&\cmark&$74\pm77$&$0.080\pm0.103$\\
\textit{SIafterVSI}&0.5&$a=3~\cm$&$10$, $320$&\cmark&$346\pm286$&$0.101\pm0.103$\\
\textit{SIafterVSI}&1&$\St=0.01$&$10$, $320$&\xmark&$2.1\pm0.6$&-\\
\textit{SIafterVSI}&1&$a=3~\mm$&$10$, $320$&\cmark&$49\pm18$&$0.001\pm0.006$\\
\textit{SIafterVSI}&1&$a=3~\mm$&$20$, $640$&\cmark&$351\pm322$&$0.053\pm0.044$\\
\textit{SIafterVSI}&1&$\St=0.1$&$10$, $320$&\cmark&$155\pm80$&$0.039\pm0.052$\\
\textit{SIafterVSI}&1&$a=3~\cm$&$10$, $320$&\cmark&$480\pm242$&$0.122\pm0.095$\\
\textit{SIafterVSI}&2&$\St=0.01$&$10$, $320$&\xmark&$15\pm10$&-\\
\textit{SIafterVSI}&2&$a=3~\mm$&$10$, $320$&\cmark&$103\pm58$&$0.016\pm0.016$\\
\textit{SIafterVSI}&2&$\St=0.1$&$10$, $320$&\cmark&$335\pm326$&$0.095\pm0.077$\\
\textit{SIafterVSI}&2&$a=3~\cm$&$10$, $320$&\cmark&$2005\pm1452$&$0.009\pm0.010$\\
\hline
\hline
\end{tabular}
}
\tablefoot{
\tablefoottext{a}{Initial dust-to-gas surface density ratio.}
\tablefoottext{b}{Given either as a size~$a$ or as a Stokes number~$\St$.}
\tablefoottext{c}{Maximum dust-to-gas volume density ratio, averaged either over strong-clumping phase or, if strong clumping does not occur, over quasi-steady phase.}
\tablefoottext{d}{Planetesimal formation rate (fraction of dust particles that becomes associated with a Roche-unstable overdensity) per unit time, averaged over strong-clumping phase.}
}
\label{table:statistics}
\end{table*}

We now discuss the dependence of the maximum dust-to-gas volume density ratio on two physical parameters, the dust size and the initial dust-to-gas surface density ratio, as well as one numerical parameter, the simulation resolution. Figure~\ref{fig:density_ratio_maximum_comparison} shows the evolution of the maximum volume density ratio in simulations of the streaming instability alone and of the scenarios \emph{SIwhileVSI} and \emph{SIafterVSI} with different combinations of these three parameters. Additionally, the average maximum during either the strong-clumping phase for all simulations that evolve into such a phase or during the quasi-steady phase for all other simulations is listed in Table~\ref{table:statistics}.

To begin with, the values of all three parameters being equal, the maximum volume density ratio is higher in the scenario \emph{SIafterVSI} than in the scenario \emph{SIwhileVSI}, and lowest in the model of only the streaming instability. This can be gathered from Table~\ref{table:statistics}, and from comparing the maximum volume density ratios in the simulations of the three scenarios with an initial surface density ratio of~$1\%$ and a dust size of~$3~\cm$ that are depicted as orange lines in Fig.~\ref{fig:density_ratio_maximum_comparison}. It underlines that the vertical shear instability and the streaming instability in concert cause stronger dust concentration than the streaming instability in isolation.

Furthermore, in agreement with previous studies of the streaming instability \citep{Bai2010b, Johansen2015, Carrera2015, Yang2017, Li2021}, we find that an increase in either the initial surface density ratio or the dust size generally results in a larger maximum volume density ratio. In addition, it is higher if the dust size is fixed at~$3~\cm$ or~$3~\mm$, corresponding to Stokes numbers ranging from~${\sim}0.05$ or~${\sim}0.005$ at~$\mbox{r=10~\au}$ to~${\sim}0.5$ or~${\sim}0.05$ at~$\mbox{r=100~\au}$ in the mid-plane of our model (see Fig.~\ref{fig:dust_size_Stokes_number} and Eq.~\ref{eq:Stokes_number}), than if the Stokes number is fixed at~$0.1$ or~$0.01$. We attribute this to dust piling up at small radii in the former case but not in the latter one. The speed of the radial dust drift can be expressed as
\begin{equation}
    v_{\dust,r}=\frac{2\St f_{\gas}^2\Pi c_{\textrm{s}}}{1+\St^2 f_{\gas}^2},
\end{equation}
where~$f_{\gas}$ is the local ratio of gas density to total density of dust and gas \citep{Nakagawa1986}. Since~\mbox{$\Pi(z=0)\propto r^{1/4}$} (see Eq.~\ref{eq:pressure_gradient}) and~\mbox{$c_{\textrm{s}}\propto r^{-1/4}$} in our model, the drift speed is independent of the radius in our simulations with a fixed Stokes number (if~$f_{\gas}$ is radially constant). On the other hand, in simulations with a fixed dust size the Stokes number increases with the radius, and thus also the drift speed. 

An enhanced resolution also entails an increase in the maximum volume density ratio, both during the quasi-steady phase and while strong clumping occurs. The former is demonstrated by our model of the streaming instability with an initial surface density ratio of~$1\%$ and a Stokes number of~$0.1$ which is shown in the left panel of Fig.~\ref{fig:density_ratio_maximum_comparison}, with the maximum volume density ratio being greater by a factor of a few if the resolution is twice as high. Similar increases are found by \citet[][see their Fig.~1]{Johansen2015} and \citet{Yang2017}. On the other hand, doubling the resolution in the model of the scenario \emph{SIafterVSI} with the same initial surface density ratio but a dust size of~$3~\mm$ that is depicted in the right panel of the figure results in an enhancement of the maximum volume density ratio by more than an order of magnitude during the strong-clumping phase. This is again in agreement with the enhancement in previously presented models of the streaming instability \citep{Yang2014, Johansen2015, Yang2017}.   

Neither our simulation of the streaming instability with the fiducial resolution nor the one with doubled resolution develops into a strong-clumping phase, so the increased resolution does not lead to strong clumping by itself. While this is consistent with the findings by \citet{Li2021}, \citet{Yang2014} and \citet{Yang2017} show that enhancing the resolution can indeed cause such a transition to strong clumping.\footnote{On the other hand, \citet{Bai2010b} find strong clumping in one of their three-dimensional simulations with a reduced resolution but not in the corresponding fiducial-resolution simulation.} It is thus important to keep in mind that whether or not strong clumping and potentially planetesimal formation arise in models of the streaming instability (and the vertical shear instability) can be dependent on the numerical resolution.

\subsection{Mid-plane dust-to-gas density ratio}
\begin{figure}[t]
  \centering
  \includegraphics[width=\columnwidth]{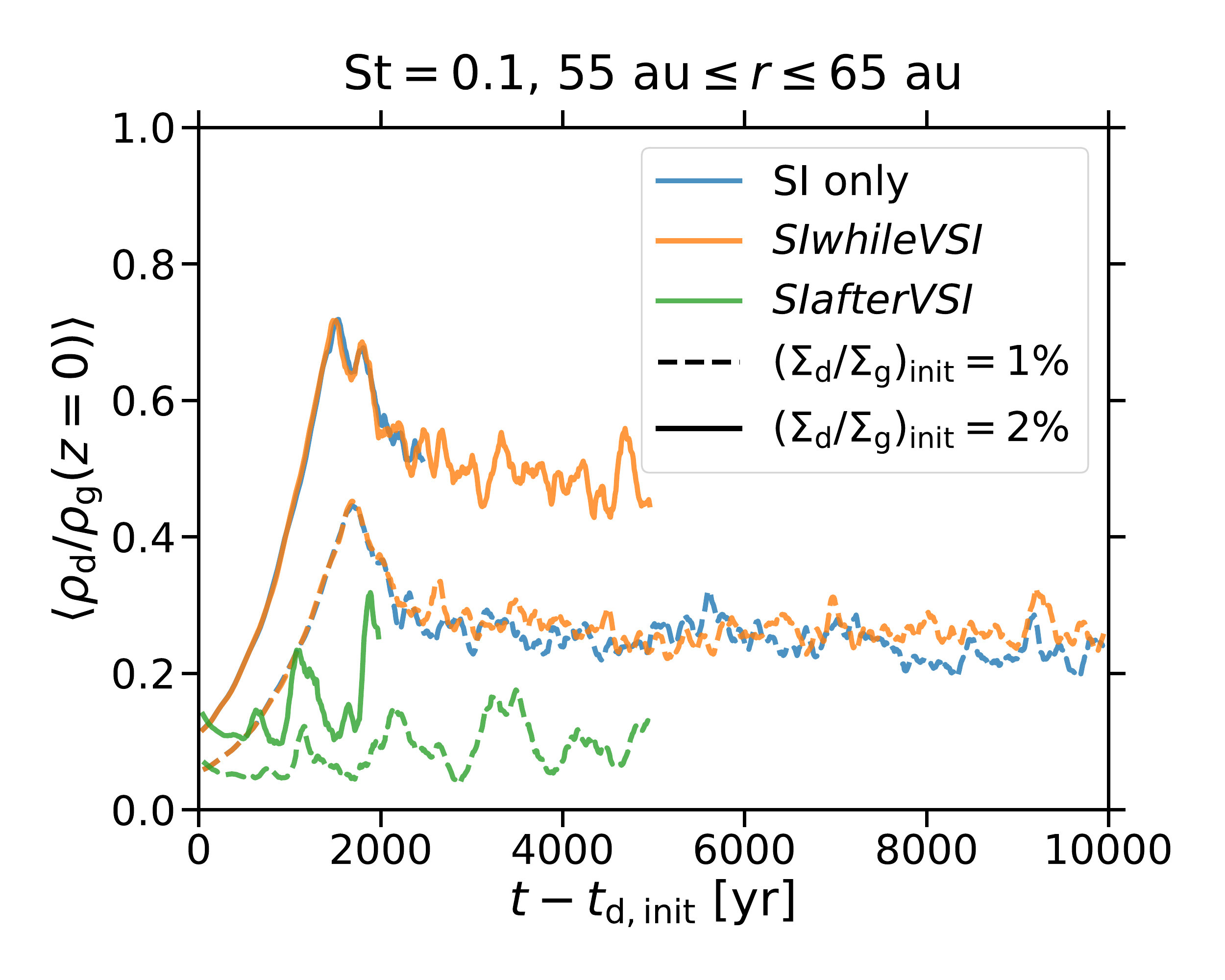}
  \caption{Dust-to-gas density ratio~$\rho_{\dust}/\rho_{\gas}$ in the mid-plane as a function of time after the dust particle initialisation~$t-t_{\textrm{d,init}}$. The mid-plane density ratio is computed as the mean over~\mbox{$r=55~\au$} to~$65~\au$ and the rolling average over~$100~\yr$. Simulations of the streaming instability only as well as the scenarios \emph{SIwhileVSI} and \emph{SIafterVSI} with initial dust-to-gas surface density ratios of~$1\%$ (dashed lines) or~$2\%$ (solid lines) and a Stokes number of~$0.1$ are shown. The mid-plane density ratio is similar in the simulations of the streaming instability and the scenario \emph{SIwhileVSI}, and lower in the scenario \emph{SIafterVSI}. While the ratio fluctuates throughout the latter simulation, it initially attains a maximum before evolving into a quasi-steady state in the former two. In all three cases, doubling the initial surface density ratio entails an increase by about a factor of two also in the mid-plane density ratio.}
  \label{fig:mid-plane_density_ratio}
\end{figure}

As noted in Sect.~\ref{sect:dust_distribution}, dust overdensities are spread out over the entire vertical extent of the dust layer. Therefore, even though local maxima in the dust-to-gas volume density ratio are of the order of ten or a hundred (see Fig.~\ref{fig:density_ratio_maximum_mean}), the mid-plane density ratio in general remains less than one. Figure~\ref{fig:mid-plane_density_ratio} shows the evolution of the mid-plane density ratio, averaged over the range of radii depicted in Fig.~\ref{fig:density_ratio}, in the same simulations as can be seen in that figure as well as in the corresponding simulations with an initial dust-to-gas surface density ratio of~$2\%$.

We find the mid-plane density ratio not to be indicative of dust concentration in the scenario \emph{SIafterVSI}. It is lowest in this scenario, whereas dust overdensities are greatest. In addition, the fluctuations of the mid-plane density ratio are likely the result of oscillations of the wave-shaped dust layer rather than the building up and breaking up of dust overdensities. The mid-plane density ratio increases with the dust-to-gas surface density ratio because dust-induced buoyancy increasingly suppresses the vertical shear instability (\citealt{Lin2019}; \citetalias{Schafer2020}), which is the main driver of turbulence in the dust layer in this scenario.

Compared with the scenario \emph{SIafterVSI}, the dust scale height is smaller (see Sect.~\ref{sect:dust_distribution}) and the mid-plane density ratio therefore higher in the model of the streaming instability in isolation and in the scenario \emph{SIwhileVSI}. It is comparable in these two models since in both of them the turbulence in the dust layer is predominantly caused by the streaming instability.

Nonetheless, as in the case of the scenario \emph{SIafterVSI}, the mid-plane density ratio does not reflect dust concentration in the model of the streaming instability alone and in the scenario \emph{SIwhileVSI}. This can be gathered from the fact that the mid-plane density ratio reaches its maximum early owing to sedimentation and remains close to constant at a lower value afterwards, while maxima in the dust-to-gas surface density ratio (see Fig.~\ref{fig:surface_density_azimuthal_velocity}) as well as in the maximum dust-to-gas volume density ratio (see Fig.~\ref{fig:density_ratio_maximum_comparison}) are attained later. The simulations of the streaming instability presented by \citet[][see their Figs.~2 and~3]{Flock2021} exhibit a similar discrepancy between the evolution of the mid-plane dust-to-gas mass ratio and the maximum dust-to-gas mass ratio as well as the dust surface density.

Furthermore, the mid-plane density ratio is almost exactly twice as high if the initial surface density ratio is doubled from~$1\%$ to~$2\%$, while the maximum volume density ratio increases non-linearly for these surface density ratios \citep[see Fig.~\ref{fig:density_ratio_maximum_comparison};][]{Johansen2009, Johansen2015}. In the streaming instability simulations by \citet{Li2021}, the mid-plane density ratio as well increases only approximately linearly with the surface density ratio, while the maximum volume density ratio can be enhanced by orders of magnitude.

\subsection{Fraction of dust mass in Roche-unstable overdensities}
\begin{figure*}[t]
  \centering
  \includegraphics[width=0.8\textwidth]{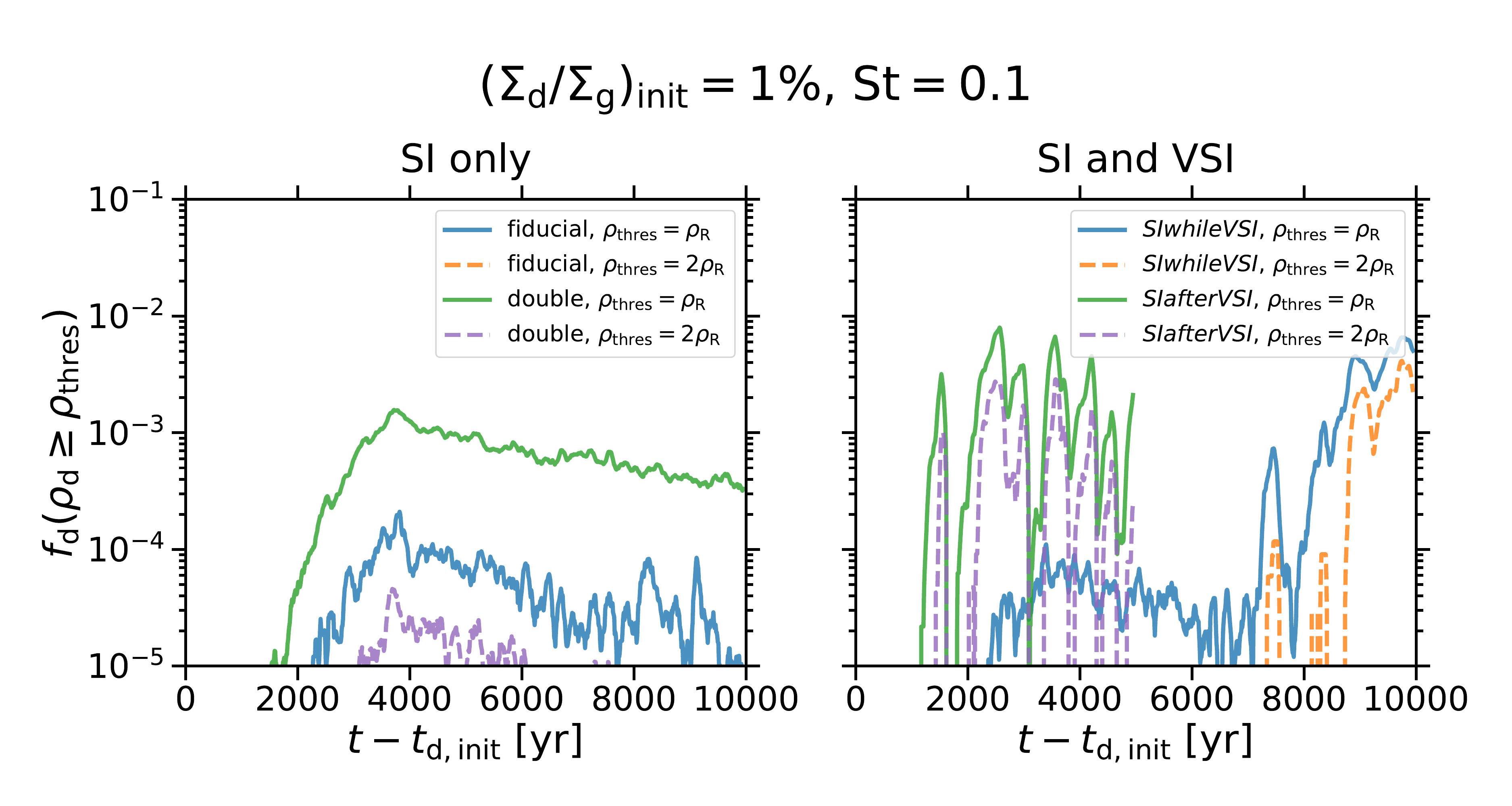}
  \caption{Fraction of total dust mass~$f_{\dust}$ that is located where the dust density~$\rho_{\dust}$ is greater than a threshold density~$\rho_{\textrm{thres}}$ as a function of time after the dust particle initialisation~$t-t_{\textrm{d,init}}$. The fraction is calculated as the rolling average over~$100~\yr$. In addition to the simulations shown in Figs.~\ref{fig:surface_density_azimuthal_velocity} and~\ref{fig:density_ratio_maximum_mean}, an analogous simulation of the streaming instability with doubled resolution is depicted. We choose threshold densities equal to once (solid lines) and twice the Roche density (dashed lines). In both the \emph{SIwhileVSI} and the \emph{SIafterVSI} simulation, Roche-unstable overdensities temporarily comprise as much as~$1\%$ of the dust mass, and overdensities that exceed twice the Roche density a slightly smaller fraction. Nevertheless, the fraction fluctuates by an order of magnitude or more over time. In contrast, in the simulations of the streaming instability with the doubled and the fiducial resolution only~$0.1~\%$ and $0.01\%$, respectively, of the dust mass is part of overdensities greater than the Roche density, and a fraction that is less by at least an order of magnitude is contained in overdensities of twice the Roche density or more.}
  \label{fig:fraction_Roche_unstable}
\end{figure*}

In Section~\ref{sect:maximum_density_ratio}, we show that the maximum dust-to-gas volume density ratio is higher in models which undergo a strong-clumping phase than in models which do not. However, it remains to be investigated whether dust overdensities could undergo gravitational collapse and form planetesimals if self-gravity were included in either or both kinds of models. To this end, in Fig.~\ref{fig:fraction_Roche_unstable} we show the fraction of the total dust mass that is associated with overdensities of at least once or twice the Roche density. Our simulations of only the streaming instability and of the scenarios \emph{SIwhileVSI} and \emph{SIafterVSI} that can be seen also in Figs.~\ref{fig:density_ratio},~\ref{fig:surface_density_azimuthal_velocity} and~\ref{fig:density_ratio_maximum_mean} as well as a simulation of the streaming instability with the same initial dust-to-gas surface density ratio and Stokes number but with doubled resolution are depicted.

The figure demonstrates that overdensities are robustly gravitationally unstable in models that experience strong clumping, but not in ones that remain in a quasi-steady state. During the strong-clumping phase in the simulations of the scenarios \emph{SIafterVSI} and \emph{SIwhileVSI}, Roche-unstable overdensities comprise up to~$1\%$ of the dust mass, and overdensities which are twice as dense a fraction that is only marginally less. The formation and dissolution of these overdensities induces variations of the fraction that are as large as an order of magnitude or more especially in the \emph{SIafterVSI} simulation, though.

In comparison, a considerably smaller fraction of about~$0.01\%$ and~$0.1\%$, respectively, of the dust mass is part of overdensities that barely exceed the Roche density in the streaming-instability-only simulations with the fiducial resolution and with the doubled resolution, which both do not undergo strong clumping (see Fig.~\ref{fig:density_ratio_maximum_comparison}). On top of that, the fraction that is contained in overdensities of at least twice the Roche density is less by an order of magnitude or more in these two simulations.

\subsection{Planetesimal formation rate}
\begin{figure}[t]
  \centering
  \includegraphics[width=\columnwidth]{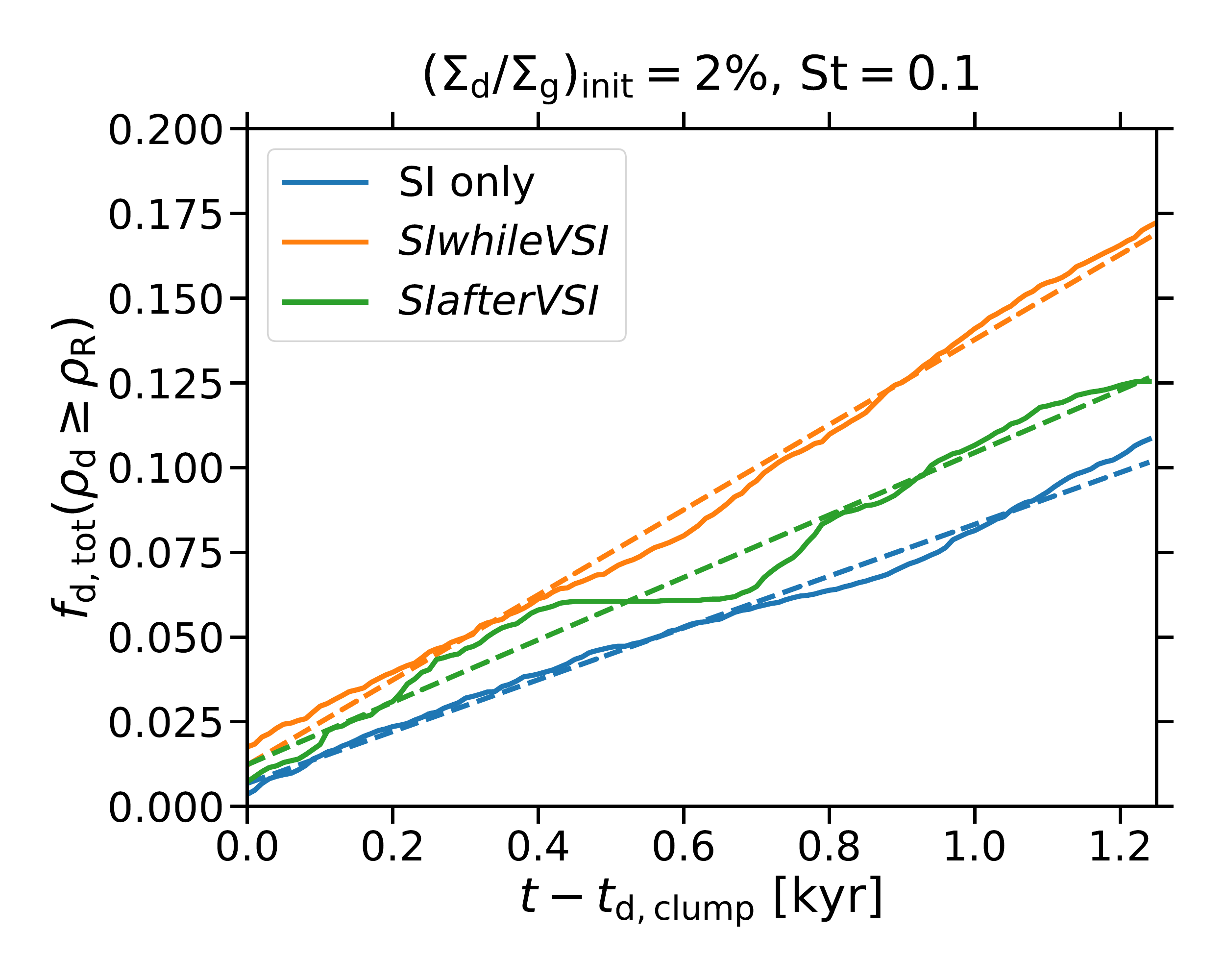}
  \caption{Cumulative fraction of the total dust mass~$f_{\textrm{d,tot}}$ that has encountered a location where the dust density~$\rho_{\dust}$ exceeds the Roche density~$\rho_{\textrm{R}}$ as a function of the time~$t$ after the start of the strong-clumping phase~$t_{\textrm{d,clump}}$. Simulations of the streaming instability in isolation (blue lines) as well as of the scenarios \emph{SIwhileVSI} (orange lines) and \emph{SIafterVSI} (green lines) with an initial dust-to-gas surface density ratio of~$2\%$ and a Stokes number of~$0.1$ are depicted. The cumulative fractions (solid lines) increase close to linearly (dashed lines) with time, with the increase being strongest in the scenario \emph{SIwhileVSI} and weakest if only the streaming instability is taken into account.}
  \label{fig:fraction_Roche-unstable_cumulative}
\end{figure}

Assuming that every overdensity that exceeds the Roche density collapses in its entirety to form one or multiple planetesimals, the fraction of the total dust mass that is part of such overdensities at a given time is equivalent to a momentary planetesimal formation efficiency. However, since gravitational collapse and planetesimal formation are not actually included in our model, each dust particle can be part of multiple such overdensities during the course of the simulations.

We therefore randomly select~$10\,000$ dust particles at the start of the strong-clumping phase in each simulation which develops into such a phase, and for each of these particles track how much time passes until the dust density at its momentary location for the first time is greater than the Roche density. Figure~\ref{fig:fraction_Roche-unstable_cumulative} shows the cumulative fraction of the total dust mass that has been part of a Roche-unstable overdensity in our simulations of the streaming instability only as well as the scenarios \emph{SIwhileVSI} and \emph{SIafterVSI} with an initial dust-to-gas surface density ratio of~$2\%$ and a Stokes number of~$0.1$. Since we generally find a roughly linear increase of this cumulative fraction with time, we compute average fractions that become associated with a Roche-unstable overdensity per unit time. We refer to these as planetesimal formation rate, with a rate of~$1~\kyr^{-1}$ indicating that all particle mass would be converted to planetesimal mass within~$1~\kyr$. For all simulations in which strong clumping occurs, we list these planetesimal formation rates in Table~\ref{table:statistics}.

Analogous to the trends we describe for the maximum dust-to-gas volume density ratio in Sect.~\ref{sect:maximum_density_ratio}, we find these rates to be higher if any of the initial surface density ratio, the initial dust size or Stokes number, and the resolution are greater. There is one notable difference, though, which can be gathered both from the figure and the table: While the maximum volume density ratio is greatest in the scenario \emph{SIafterVSI}, the rates are by tendency largest in the scenario \emph{SIwhileVSI} -- the overall highest rate is attained in a simulation of the streaming instability alone, though. The reason for this discrepancy probably lies in dust diffusion being stronger in the scenario \emph{SIafterVSI}, where the vertical shear instability is the main source of turbulence in the dust layer, than in the scenario \emph{SIwhileVSI} or in our model of the streaming instability in isolation, where it is (predominantly) caused by this instability. Overdensities are thus more prone to dispersing and forming anew. This explanation is supported by a comparison of the upper panels of Fig.~\ref{fig:surface_density_azimuthal_velocity} as well as both by the fluctuations in the fraction of the total dust mass that is comprised in Roche-unstable overdensities being greatest in the scenario \emph{SIafterVSI}, as can be seen from Fig.~\ref{fig:fraction_Roche_unstable}, and by the standard deviations of the planetesimal formation rates being comparatively large in this scenario.

\section{Thresholds for planetesimal formation}
\label{sect:thresholds}
\begin{figure}[t]
  \centering
  \includegraphics[width=\columnwidth]{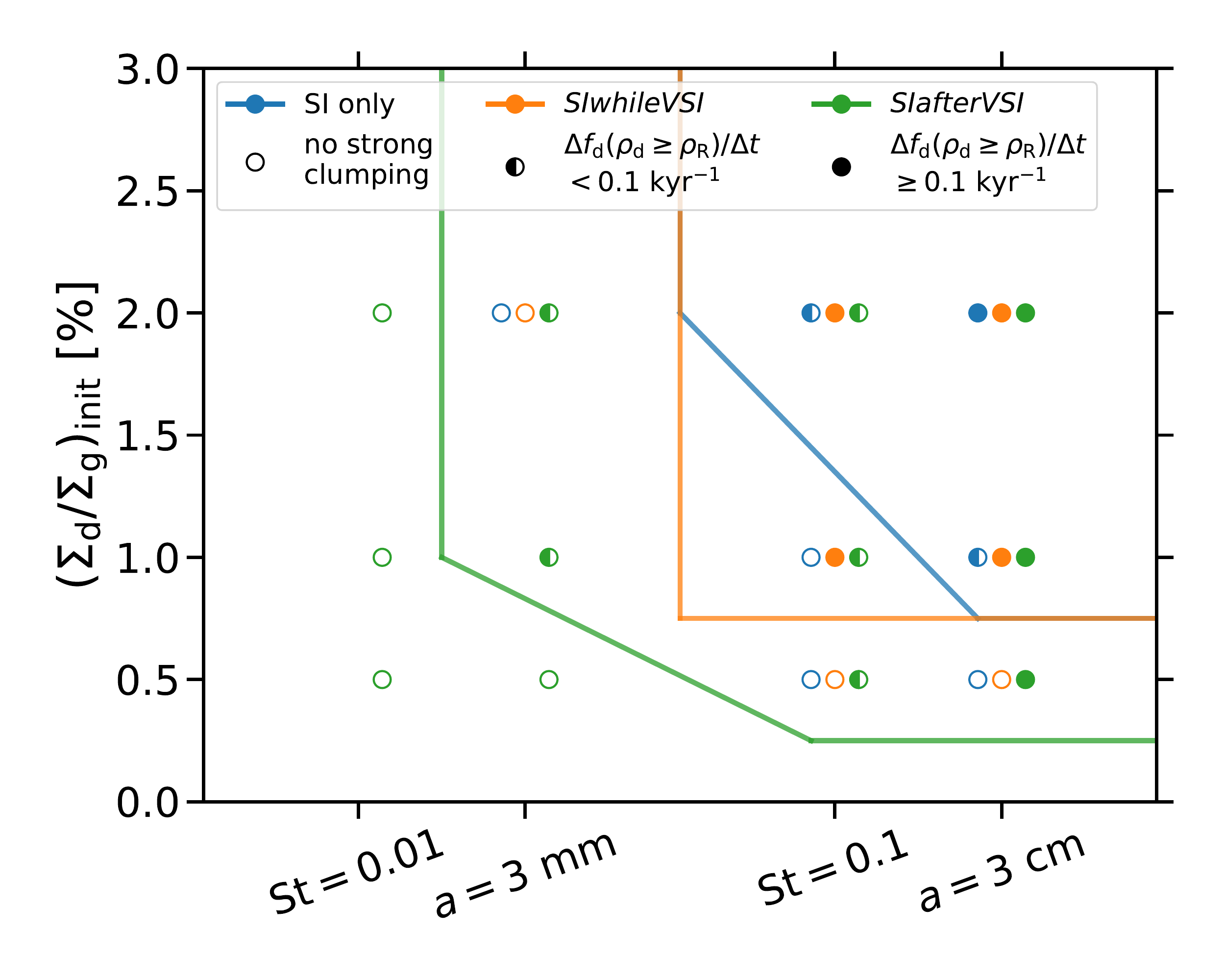}
  \caption{Thresholds for strong clumping as well as planetesimal formation rates. Initial dust-to-gas surface density ratios~$(\Sigma_{\dust}/\Sigma_{\gas})_{\textrm{init}}$ and Stokes numbers~$\St$ or dust sizes~$a$ are plotted on the ordinate and the abscissa, respectively. Blue, orange, and green colors represent our models of the streaming instability in isolation, the scenario \emph{SIwhileVSI} and the scenario \emph{SIafterVSI}. Simulations that do not experience strong clumping are depicted as empty circles; ones in which the planetesimal formation rate during the strong-clumping phase is less than~$0.1~\kyr^{-1}$ are represented by half-filled circles and ones in which the rate is greater than this value as filled circles. Lines separate combinations of (surface density ratio, dust size or Stokes number) for which strong clumping does or does not occur. The minimum values of these two parameters required for strong clumping are smallest in the scenario \emph{SIafterVSI}, with the combinations~($0.5\%$,~\mbox{$\St=0.1$}) and~($1\%$,~\mbox{$a=3~\mm$}) being sufficient. The planetesimal formation rates, on the other hand, are by tendency highest in the scenario~\emph{SIwhileVSI}.}
  \label{fig:thresholds}
\end{figure}

Based on what we discussed in the previous sections, we conclude that if strong clumping occurs in a simulation of ours, then planetesimals would form in the simulation if self-gravity were included. This is since overdensities in simulations that undergo strong clumping robustly exceed the Roche density. Whether a simulation develops into a strong-clumping phase is evident, for instance, from the evolution of the maximum dust-to-gas volume density ratio. The maximum is well-correlated with the mean and standard deviation of the volume density ratio and thus representative of dust concentration at large. From Figure~\ref{fig:thresholds} and Table~\ref{table:statistics}, it can be gathered in which of our simulations strong clumping and thus potentially planetesimal formation occur.

Overall, we find that the vertical shear instability and the streaming instability together cause stronger dust concentration than the streaming instability in isolation. This is reflected in these two instabilities inducing strong clumping for combinations of (initial dust-to-gas surface density ratio, dust size or Stokes number) as small as~($0.5\%$,~\mbox{$\St=0.1$}) or~($1\%$,~\mbox{$a=3~\mm$}) in the scenario~\emph{SIafterVSI}, while in the scenario~\emph{SIwhileVSI} at least~($1\%$,~\mbox{$\St=0.1$}) is required. In comparison, the thresholds are higher when only the streaming instability is considered, with~($1\%$,~\mbox{$a=3~\cm$}) or~($2\%$,~\mbox{$\St=0.1$}) being necessary. 

In agreement with our findings, planetesimal formation arises for~($1\%$,~\mbox{$\St\approx0.1$}), but not for~($1\%$,~\mbox{$\St\approx0.01$}), in the models of the streaming instability including a prescribed pressure bump that are presented by \citet{Carrera2021} and \citet{Carrera2022b}. On the other hand, \citet{Flock2021} and \citet{Li2021} find the streaming instability to lead to planetesimal formation for~($1\%$,~\mbox{$\St=0.1$}) also in the absence of a pressure bump.

In addition to a differentiation between simulations which do or do not experience strong clumping, Fig.~\ref{fig:thresholds} depicts trends in the planetesimal formation rate. Half-filled circles represent simulations in which the rate does not exceed~$0.1~\kyr^{-1}$, that is to say after~$1~\kyr$ less than~$10\%$ of the total dust mass would have been converted to planetesimal mass, while filled circles represent simulations with a rate equal to or greater than this threshold value. As we discuss in the previous section, the rates are generally highest in the scenario \emph{SIwhileVSI} and exceed the threshold in all cases in this scenario. In the scenario \emph{SIafterVSI}, on the other hand, the rates are greater than the threshold only in simulations with dust size of~$3~\cm$, and in our model of the streaming instability in isolation only for this dust size and an initial dust-to-gas surface density ratio of~$2\%$. 

\section{Discussion}
\label{sect:discussion}
\subsection{General implications}
We study dust concentration and the potential for planetesimal formation in two-dimensional global simulations of the streaming instability and the vertical shear instability. These simulations represent three different scenarios, two in which the two instabilities coexist and one involving only the streaming instability, and include different initial dust-to-gas surface density ratios, dust sizes or Stokes numbers, respectively, and resolutions.

In the scenario \emph{SIafterVSI}, the vertical shear instability attains a saturated state before the streaming instability begins to grow, while in the scenario \emph{SIwhileVSI} both instabilities start to develop at the same time. Both scenarios are pausible since the growth from micron-sized dust grains to the millimetre- or centimetre-sized dust aggregates that we simulated takes thousands of orbital periods \citep{Zsom2010, Lorek2018}, but it is unclear at which point during this growth protoplanetary disks have evolved into a state similar to the one in our model which is favourable to the development of the vertical shear instability.

Our simulations show that the vertical shear instability causes the formation of gas pressure bumps that promote dust concentration by the streaming instability. This is consistent with both the findings that the vertical shear instability induces dust concentration in pressure bumps \citep{Stoll2016} and that pressure bumps -- be they prescribed \citep{Lenz2019, Carrera2021, Carrera2022a, Carrera2022b, Lehmann2022, Xu2022b} or caused by the magnetorotational instability under the assumption of ideal magnetohydrodynamics \citep{Johansen2007b, Johansen2011} -- facilitate dust accumulation and planetesimal formation owing to the streaming instability. The same has been found for vortices induced by the vertical shear instability \citep{Lehmann2022} and by the subcritical baroclinic instability \citep{Raettig2015, Raettig2021}.

We note that \citet{Xu2022a} present simulations of the magnetorotational instability including ambipolar diffusion in which the instability gives rise to pressure bumps where dust accumulates. In contrast to in our model, though, they find that this dust accumulation is a consequence of the mutual drag between gas and dust, but not of the streaming instability, because it occurs in pressure maxima where there is no pressure gradient to drive the streaming instability \citep[but see][]{Auffinger2018, Lin2022}.

A key result of our study is that dust overdensities are sufficient to form planetesimals for lower dust-to-gas surface density ratios and smaller dust sizes if both the vertical shear instability and the streaming instability are considered than when only the streaming instability is taken into account. Both our and previous work \citep{Carrera2015, Yang2017, Li2021} has shown that either dust-to-gas ratios that are greater than the canonical interstellar medium value of~$1\%$ or dust sizes which are larger than what is observed in protoplanetary disks are required for the streaming instability in isolation to lead to planetesimal formation.

On the other hand, in our scenario \emph{SIafterVSI} -- in which the vertical shear instability has saturated before the streaming instability begins to grow -- we find planetesimal formation to be possible for a surface density ratio of~$1\%$ and a dust size of~$3~\mm$, in agreement with the sizes derived from observed opacity spectral indices \citep[e.g.][]{Sierra2019, Macias2019, Macias2021, Carrasco-Gonzalez2019}, though larger than the ones inferred from polarisation measurements \citep[e.g.][]{Ohashi2020, Mori2021a}. That is, if this scenario applies to some or all protoplanetary disks, planetesimal formation via the vertical shear instability and the streaming instability should be omnipresent in the parts of these disks where the vertical shear instability is active, which roughly correspond to the region between~$10~\au$ and~$100~\au$ covered by our simulation domains \citep{Lin2015, Malygin2017, Pfeil2019}.

At first glance, our results seem to contradict previous studies finding that turbulence reduces the growth rate of the linear streaming instability \citep{Umurhan2020, Chen2020} and that driven Kolmogorov-like turbulence inhibits planetesimal formation owing to the non-linear instability \citep{Gole2020}. We note, though, that in these studies turbulence is purely a source of isotropic diffusion and viscosity, while our and the above-mentioned work evinces that instabilities which drive turbulence also give rise to pressure bumps and vortices. In addition, turbulence is not generally isotropic, and purely vertical diffusion is not detrimental to radial dust concentration (\citealt{Yang2018}; \citetalias{Schafer2020}).

\subsection{Implications for one-dimensional models including planetesimal formation}
In one-dimensional models of protoplanetary disks that include a prescription of the formation of planetesimals via the streaming instability, it is often assumed that the mid-plane dust-to-gas density ratio needs to exceed unity for planetesimal formation to occur \citep[e.g.][]{Drazkowska2014, Drazkowska2016, Drazkowska2017, Schoonenberg2017, Schoonenberg2018, Stammler2019}. This condition is based on the simulations by \citet{Johansen2007a}, which notably do not include the vertical stellar gravity.

However, while this condition might be sufficient, our model shows that it is not necessary. We find that dust concentration is sufficiently strong for planetesimal formation in all of our simulations with an initial dust-to-gas surface density ratio of~$2\%$ and a Stokes number of~$0.1$, particularly also in the one of the streaming instability alone (see Fig.~\ref{fig:thresholds}). Nevertheless, it can be seen from Fig.~\ref{fig:mid-plane_density_ratio} that the mid-plane density ratio remains less than one in these simulations.

We therefore recommend to base prescriptions of planetesimal formation owing to the streaming instability on whether the dust-to-gas surface density ratio and the dust size exceed the threshold values given by our Fig.~\ref{fig:thresholds} if both the streaming instability and the vertical shear instability are modeled; and by this figure, by Eqs.~8 and~9 of \citet{Yang2017}, or by Eq.~10 of \citet{Li2021} if only the streaming instability is taken into consideration. If planetesimal formation is established to occur, our Table~\ref{table:statistics} lists planetesimal formation rates, rates at which dust mass is converted to planetesimal mass, for a given combination of surface density ratio and dust size in each of our three scenarios.

\subsection{Limitations}
While we investigate which conditions, in terms of dust-to-gas surface density ratio and dust size, are necessary for the vertical shear instability and streaming instability in combination or the streaming instability in isolation to induce planetesimal formation, we can not actually model this process since our two-dimensional simulations do not include self-gravity. As with previous similar parameter studies \citep{Carrera2015, Yang2017, Li2021}, this is because it is not computationally feasible to cover a significant fraction of the parameter space with three-dimensional simulations.

Nonetheless, we discuss in detail in Sect.~\ref{sect:metrics} which metrics are applicable to two-dimensional models to gauge the potential for planetesimal formation in equivalent three-dimensional ones. Furthermore, it has been shown that the surface density ratios and dust sizes that are necessary for dust concentration to be sufficient for planetesimal formation are comparable in two- and three-dimensional simulations of the streaming instability \citep{Yang2017, Li2021} and vertical shear instability \citep{Lehmann2022}. Nevertheless, the non-linear regimes of the streaming instability \citep{Kowalik2013} and the vertical shear instability \citep{Nelson2013, Stoll2014}, in contrast to their linear regimes \citep{Youdin2005, Nelson2013, Barker2015}, are not axisymmetric. In particular, the vertical shear instability gives rise to vortices in which dust accumulates \citep{Flock2020, Lehmann2022}.

More generally, our model could be improved upon by deviating from the assumption of a constant initial surface density ratio and dust size or Stokes number, and instead considering the structures of rings and gaps with varying dust and gas surface densities and maximum dust sizes that are observed in protoplanetary disks \citep[e.g.][]{Macias2019, Macias2021, Carrasco-Gonzalez2019, Andrews2020}. This could entail a need for other parameters than the surface density ratio and dust size, for instance the dust flux \citep{Lenz2019, Flock2021}, to describe the conditions that are required for planetesimal formation to occur. We further do not consider dust size distributions \citep{Bai2010b, Schaffer2018, Schaffer2021, Krapp2019, Zhu2021, Yang2021d}. In addition, the gas disk is in our model is only affected by the stellar gravity as well as the vertical shear instability and the streaming instability, processes like disk winds and other instabilities are not taken into account; we simulated a simple isothermal or adiabatic equation of state rather than the various disk heating and cooling processes; and we do not consider dust and gas chemistry.

\section{Summary}
\label{sect:summary}
We employ two-dimensional, axisymmetric adaptive mesh refinement simulations of the vertical shear instability and the streaming instability, which cover the outer regions of protoplanetary disks on a global scale, to identify the threshold values of dust-to-gas surface density ratio and dust size or Stokes number that are required for the two instabilities in conjunction or the streaming instability in isolation to induce planetesimal formation. Similar parameter studies are presented by \citet{Carrera2015}, \citet{Yang2017}, and \citet{Li2021}, though all of these are based on two-dimensional local shearing box simulations of the streaming instability only.

Since self-gravity is not included in our two-dimensional simulations, planetesimal formation is not actually modeled. We therefore dedicate a detailed discussion to metrics that can be applied to differentiate between simulations in which it would occur and ones in which it would not:
\begin{itemize}
    \item The maximum dust-to-gas volume density ratio is well-correlated with the mean and standard deviation of the volume density ratio and thus representative of the strength of dust concentration at large.
    \item The maximum volume density ratio increases if either the initial dust-to-gas surface density ratio, the dust size, or the resolution is enhanced. This is in agreement with previous work on the streaming instability \citep{Bai2010b, Yang2014, Johansen2015, Carrera2015, Yang2017, Li2021}. Moreover, it is higher in our models with a fixed dust size than in ones with a comparable fixed Stokes number. This is because the drift speed increases with the radial distance to the star in the former case (but not in the latter), resulting in dust piling up in the radial dimension.
    \item While some simulations remain in a quasi-steady state after the dust has settled to a mid-plane layer, others evolve into a phase of strong dust clumping \citep{Johansen2015, Yang2017, Li2021}. This phase is characterised by comparatively high maxima and large fluctuations in the maximum volume density ratio, which are a consequence of the formation and dissolution of strong overdensities. While whether or not strong clumping occurs does not depend on the resolution in our study, a transition to strong clumping is found in previous studies of the streaming instability if the resolution is increased \citep{Yang2014, Yang2017}.
    \item Only in simulations that undergo a strong-clumping phase are overdensities robustly Roche-unstable, that is to say only in these simulations is a significant fraction of the total dust mass comprised in overdensities that exceed once and even twice the Roche density.
    \item The mid-plane dust-to-gas density ratio is not suitable as an indicator of local dust concentration because local maxima of the volume density ratio can be found at all heights within the dust layer.
\end{itemize}

This leads us to the conclusion that models in which planetesimal formation would be possible if self-gravity were taken into account distinguish themselves from ones in which it would not in that they develop into a strong-clumping phase. From Figure~\ref{fig:thresholds} and Table~\ref{table:statistics}, it can be gathered which of our simulations do or do not experience strong clumping. Consistent with the trends described above for the maximum volume density ratio, these are simulations with high initial surface density ratios and large dust sizes or Stokes numbers. 

Importantly, we find that the minimum surface density ratios and dust sizes or Stokes numbers which are necessary for strong clumping and thus potentially planetesimal formation are lower in our models of both the vertical shear instability and the streaming instability than in our model of only the latter. They are smallest if the vertical shear instability has saturated before the streaming instability begins its growth, and higher if both instabilities start their growth at the same time. The reason for this lies in the vertical shear instability giving rise to pressure bumps in which dust accumulates, with these accumulations in turn seeding further dust concentration by the streaming instability.

Since the cumulative fraction of the dust mass that has been part of a Roche-unstable overdensity increases largely linearly with time in our simulations, we calculate planetesimal formation rates. These rates are listed in Table~\ref{table:statistics} and indicate which fraction of the dust mass becomes associated with such an overdensity and would be converted to planetesimal mass per unit time. They exceed~$0.1~\kyr^{-1}$ in several cases and~$0.25~\kyr^{-1}$ in the best case, and are again by tendency greater in our models of both instabilities in concert than in our model of the streaming instability in isolation, though largest if both instabilities grow simultaneously.

\begin{acknowledgements}
We are thankful to the anonymous referee for their comments that helped to improve in particular the clarity of this paper. To analyse and visualise the simulations presented in this paper, the Python packages yt\footnote{\url{http://yt-project.org}} \citep{Turk2011}, Matplotlib\footnote{\url{https://matplotlib.org}} \citep{Hunter2007}, and NumPy\footnote{\url{https://numpy.org}} \citep{Oliphant2006} were used. The FLASH Code has in part been developed by the DOE NNSA-ASC OASCR Flash Center at the University of Chicago. Computational resources employed to conduct the simulations were provided by the Regionales Rechenzentrum at the University of Hamburg, by the Norddeutscher Verbund für Hoch- und H\"ochstleistungsrechnen (HLRN), and by the SCIENCE HPC Center at the University of Copenhagen. U.S. and A.J. are thankful for funding from the European Research Foundation (ERC Consolidator Grant 724687-PLANETESYS). A.J. further gratefully acknowledges funding from the Knut and Alice Wallenberg Foundation (Wallenberg
Scholar Grant 2019.0442), the Swedish Research Council (Project Grant 2018-04867), the Danish National Research Foundation (DNRF Chair Grant DNRF159), and the G\"oran Gustafsson Foundation.
\end{acknowledgements}

\bibliography{Planetesimal_formation_thresholds_explored_in_two-dimensional_global_models}

\end{document}